\documentclass{ieeeaccess}
\usepackage{cite}
\usepackage{amsmath,amssymb,amsfonts}
\usepackage{graphicx}
\usepackage{textcomp}
\usepackage{soul}
\usepackage{cite}
\usepackage{amsmath,amssymb,amsfonts}
\usepackage{algorithmic}
\usepackage{graphicx,color}
\usepackage{textcomp}
\usepackage{multirow}
\usepackage{xcolor}
\usepackage{booktabs}
\usepackage{array}
\usepackage{hyperref}
\usepackage{threeparttable}
\usepackage{pifont} 
\usepackage{colortbl} 
\usepackage{array,soul}    

\usepackage{enumitem}
\usepackage{tabularx}
\newcommand{\cmark}{\ding{51}}%
\newcommand{\xmark}{\ding{55}}%

\usepackage{tabularx}
\hypersetup{hidelinks=true}
\usepackage{algorithm,algorithmic}

\def\BibTeX{{\rm B\kern-.05em{\sc i\kern-.025em b}\kern-.08em
    T\kern-.1667em\lower.7ex\hbox{E}\kern-.125emX}}
\AtBeginDocument{\definecolor{tmlcncolor}{cmyk}{0.93,0.59,0.15,0.02}\definecolor{NavyBlue}{RGB}{0,86,125}}

\def\authorrefmark#1{\ensuremath{^{\textbf{#1}}}}

\usepackage{bm}
\makeatletter
\AtBeginDocument{\DeclareMathVersion{bold}
\SetSymbolFont{operators}{bold}{T1}{times}{b}{n}
\SetSymbolFont{NewLetters}{bold}{T1}{times}{b}{it}
\SetMathAlphabet{\mathrm}{bold}{T1}{times}{b}{n}
\SetMathAlphabet{\mathit}{bold}{T1}{times}{b}{it}
\SetMathAlphabet{\mathbf}{bold}{T1}{times}{b}{n}
\SetMathAlphabet{\mathtt}{bold}{OT1}{pcr}{b}{n}
\SetSymbolFont{symbols}{bold}{OMS}{cmsy}{b}{n}
\renewcommand\boldmath{\@nomath\boldmath\mathversion{bold}}}
\makeatother

\def\BibTeX{{\rm B\kern-.05em{\sc i\kern-.025em b}\kern-.08em
    T\kern-.1667em\lower.7ex\hbox{E}\kern-.125emX}}

\begin{document}
\history{Date of publication xxxx 00, 0000, date of current version xxxx 00, 0000.}
\doi{10.1109/ACCESS.2024.0429000}

\title{Packet Inspection Transformer: A Self-Supervised Journey to Unseen Malware Detection with Few Samples}
\author{\uppercase{Kyle Stein}\authorrefmark{1}, \IEEEmembership{Student Member, IEEE},
\uppercase{Andrew Arash Mahyari}\authorrefmark{2}, \IEEEmembership{Member, IEEE}, \uppercase{Guillermo Francia III}\authorrefmark{3}, \IEEEmembership{Member, IEEE}, \uppercase{Eman El-Sheikh}\authorrefmark{3},
\IEEEmembership{Member, IEEE}}

\address[1]{Department of Intelligent Systems and Robotics, University of West Florida, Pensacola, FL, USA}
\address[2]{Florida Institute For Human and Machine Cognition (IHMC), Pensacola, FL, USA}
\address[3]{Center for Cybersecurity, University of West Florida, Pensacola, FL, USA}
\tfootnote{This work is partially supported by the UWF Argo Cyber Emerging Scholars (ACES) program funded by the National Science Foundation (NSF) CyberCorps® Scholarship for Service (SFS) award under grant number 1946442. Any opinions, findings, and conclusions or recommendations expressed in this document are those of the authors and do not necessarily reflect the views of the NSF.}

\markboth
{Author \headeretal: Preparation of Papers for IEEE TRANSACTIONS and JOURNALS}
{Author \headeretal: Preparation of Papers for IEEE TRANSACTIONS and JOURNALS}

\corresp{Corresponding author: Kyle Stein (ks209@students.uwf.edu).}

\begin{abstract}
As networks continue to expand and become more interconnected, the need for novel malware detection methods becomes more pronounced. Traditional security measures are increasingly inadequate against the sophistication of modern cyber attacks. Deep Packet Inspection (DPI) has been pivotal in enhancing network security, offering an in-depth analysis of network traffic that surpasses conventional monitoring techniques. DPI not only examines the metadata of network packets, but also dives into the actual content being carried within the packet payloads, providing a comprehensive view of the data flowing through networks. While the integration of advanced deep learning techniques with DPI has introduced modern methodologies into malware detection and network traffic classification, state-of-the-art supervised learning approaches are limited by their reliance on large amounts of annotated data and their inability to generalize to novel, unseen malware threats. To address these limitations, this paper leverages the recent advancements in self-supervised learning (SSL) and few-shot learning (FSL). Our proposed self-supervised approach trains a transformer via SSL to learn the embeddings of packet content, including payload, from vast amounts of unlabeled data by masking portions of packets, leading to a learned representation that generalizes to various downstream tasks. Once the representation is extracted from the packets, they are used to train a malware detection algorithm. The representation obtained from the transformer is then used to adapt the malware detector to novel types of attacks using few-shot learning approaches. Our experimental results demonstrate that our method achieves classification accuracies of up to 94.76\% on the UNSW-NB15 dataset and 83.25\% on the CIC-IoT23 dataset.
\end{abstract}


\titlepgskip=-21pt

\maketitle

\section{Introduction}

With the continuing expansion of the digital landscape, the significance of malware detection becomes increasingly important. With the modernization of sophisticated cyber attacks, traditional security measures often fall short in providing adequate protection. Traditional firewalls and antivirus software often prove inadequate since they primarily rely on predefined packet header rules through conventional packet filtering and known malware signatures, leaving them vulnerable to new and sophisticated cyber attacks that do not match existing patterns or behaviors \cite{firewall, antivirus}. Deep Packet Inspection (DPI) is a well-known tool for the analysis of network packets in the security of networks \cite{DPI1, DPI2}. DPI goes beyond conventional network monitoring tools by examining not only the five-tuple of network packets, such as source and destination Internet Protocol (IP) addresses, but also inspects the data payload within each packet. This in-depth analysis allows for a more comprehensive understanding of the nature of the traffic passing through a network, enabling the identification of malicious activities that might otherwise go unnoticed. 

With recent advances in deep learning and artificial intelligence \cite{b1,he2016deep,Attention}, the integration of these techniques into network traffic analysis and DPI has become more prevalent. The dynamic nature of cyber attacks, characterized by constantly evolving malware, necessitates a shift from traditional, signature-based detection methods to more adaptive and intelligent systems. It is estimated that 560,000 new pieces of malware are detected daily and more than 1 billion malware programs exist \cite{malwarestats}. Deep learning, with its capability to model complex patterns and adapt to evolving information through neural networks, presents a promising approach to addressing the challenges of modern malware detection. With this, cybersecurity systems can analyze data in real time, identifying threats based on behavioral patterns rather than relying solely on known signatures.

\begin{table*}[t!]
\centering
\caption{Comparative Summary of Related Works and the Proposed Approach}
\label{tab:comparative-summary}
\scriptsize  
\renewcommand{\arraystretch}{1.4} 
\setlength{\tabcolsep}{5pt} 
\rowcolors{1}{gray!10}{white} 

\begin{tabularx}{\textwidth}{|>{\raggedright\arraybackslash}p{2.3cm}|
                                 >{\raggedright\arraybackslash}p{1.8cm}|
                                 >{\raggedright\arraybackslash}p{2.2cm}|
                                 >{\raggedright\arraybackslash}p{2.7cm}|
                                 >{\centering\arraybackslash}p{2cm}|
                                 >{\raggedright\arraybackslash}X|}
\hline
\rowcolor{gray!20} 
\textbf{Reference} & 
\textbf{Data} & 
\textbf{Features} & 
\textbf{Method/Architecture} & 
\textbf{Unseen Attack?} &
\textbf{Key Observations} \\
\hline

Aceto et al. \cite{b3} &
Mobile Network Traffic &
Statistical Attributes &
Deep Learning (various) &
\xmark &
Demonstrates deep learning for mobile traffic classification; does not address unseen malware. \\
\hline

Patheja et al. \cite{Patheja} &
NSL-KDD &
Statistical Attributes &
Self-taught Learning, Softmax Regression &
\xmark &
Uses classic supervised learning with limited features; raw byte payloads not considered. \\
\hline

Doshi et al. \cite{DDOS} &
Custom IoT DDoS dataset &
Statistical Attributes &
ML Ensemble + Simple NN &
\xmark &
Focuses on DDoS detection only; lacks generalization to unseen classes. \\
\hline

Almaraz-Rivera et al. \cite{IOT} &
IoT Traffic &
Packet Counts, Bytes, Durations &
ML + DL (DDoS) &
\xmark &
Primarily scale-based features; excludes raw payload bytes and novel threats. \\
\hline

Stein et al. \cite{CANRnn} &
CAN Bus (Automotive) &
CAN Payload (8 bytes) &
RNN-based Supervised &
\xmark &
Demonstrates viability in automotive domain; payload shorter than typical IP packets. \\
\hline

Saiyeed et al. \cite{saiyed2024deep} &
Custom IoT DDoS dataset &
Traffic-Level Statistics &
Ensemble (CNN + LSTM) &
\xmark &
Excels at DDoS detection but does not target new/unseen threats. \\
\hline

He et al. \cite{transaction} &
Malware PE/APK &
RGB-Transformed Binaries &
ResNeXt w/ Attention &
\xmark &
Employs attention for malware classification; requires labeled data for known threats. \\
\hline

Zhou et al. \cite{Wf-transformer} &
Website Fingerprinting &
Packet Traces (TLS) &
Transformers (Self-Attention) &
\xmark &
Improves WF classification via attention; does not address limited labeling or novel classes. \\
\hline

Ravi \& Alazab \cite{ravi2023attention} &
Network Traffic &
Extracted Features + CNN &
Attention-based CNN &
\xmark &
Enhances classification accuracy but lacks handling of unseen malware classes. \\
\hline

Yu et al. \cite{yu2020intrusion} &
Custom Flow Data (imbalanced) &
Flow-Level Statistics &
CNN + FSL &
\xmark~ &
Alleviates class imbalance with FSL; only uses flow features, no raw payload data. \\
\hline

Lu et al. \cite{lu2023few} &
IoT Intrusions  &
Flow-Level Statistics &
CNN + FSL &
\xmark~ &
Addresses zero-day attacks; relies on high-level features, ignoring payload details. \\
\hline

Zhou et al. \cite{zhou2020siamese} &
Industrial Cyber-Physical &
CNN-Based Feature Vectors &
Siamese CNN + FSL &
\xmark~ &
Works well for anomaly detection but excludes longer packet contexts or raw bytes. \\
\hline

Stein et al. \cite{stein2024towards} &
Network and IoT Intrusions &
Raw Packet Bytes &
Supervised Transformer + FSL &
\cmark~  &
Generalizes well in FSL tasks but assumes there is enough labeled data for supervised pre-training. \\
\hline

\textbf{Proposed Method} &
Network and IoT Intrusions &
Raw Packet Bytes &
\textbf{Self-Supervised Transformer} + FSL &
\cmark~ &
Learns embeddings from raw unlabeled packet data; adapts quickly to novel/unseen attacks with few samples. \\
\hline

\end{tabularx}
\end{table*}

DPI faces several \emph{critical challenges} that limit its efficiency and effectiveness. \emph{First}, most existing methods focus primarily on statistical-based features or inspecting the few first bytes of raw packet payload data, restricting their ability to capture deeper insights. \emph{Second}, supervised learning approaches struggle to generalize beyond known attack types, as they learn representations tailored to the classes labeled in the training data \cite{b9, b10, saiyed2024deep}, resulting in poor performance when faced with novel threats or emerging attack patterns. \emph{Third}, these methods heavily rely on large amounts of annotated data, excelling in detecting malware types with abundant labeled samples but performing poorly on malware with limited or no annotations.

By leveraging few-shot learning techniques, we enable these models to adapt seamlessly to novel, unseen malware threats, setting a new standard in malware detection. The key contributions of this paper are as follows:

\begin{itemize}[leftmargin=1em]
\item \textbf{Whole Payload Inspection:} Transformers have already revolutionized natural language processing and computer vision \cite{b1}, thanks to their deep, interconnected architectures that excel at capturing long-range dependencies in sequential data. In this paper, we harness the unparalleled capabilities of transformers to redefine malware detection, moving beyond traditional methods that rely on statistical-based features and IP tracking, which are prone to false alarms and missed threats. Moreover, transformers possess the extraordinary ability to analyze entire payloads in their full context unlike convolutional neural network (CNN)-based approaches, which are limited to inspecting only the first 500 bytes of payload data. 

\item \textbf{Generalization to Unseen Malware Types:} Self-Supervised Learning (SSL) is a  paradigm that empowers models to extract deep, meaningful representations from vast amounts of unlabeled data--eliminating the need for costly and labor-intensive manual annotation \cite{b1,SSL,SSL2}. By combining the transformers' context-aware learning with the power of SSL, we train models on massive unannotated payload datasets, equipping them to decode intricate patterns within both benign and malicious payloads. This enables the model to identify complex patterns across large-scale benign and malicious payloads and achieve generalization to unseen malware types. 

\item \textbf{Learning Unseen Malware Patterns from a few Annotated Payloads:} Traditional supervised learning-based payload inspection methods rely on vast amounts of annotated data for each malware type \cite{b9, transaction, saiyed2024deep}. The challenge becomes even more severe when novel malware variants appear, as limited samples make it nearly impossible to train conventional models effectively. While transformers, pre-trained on unannotated data, offer promising generalization capabilities, they inherently lack the ability to recognize \ul{unseen} malware types. To bridge this gap, we integrate few-shot learning techniques, enabling the model to rapidly adapt to both previously encountered and entirely novel malware strains using minimal labeled examples. This approach dramatically enhances detection efficiency and accuracy, allowing AI-driven cybersecurity systems to identify and neutralize emerging threats—a crucial advancement in the fight against ever-evolving cyber attacks.

\end{itemize}

The structure of this paper is as follows: Section~\ref{sec:rw} reviews related work, while Section~\ref{sec:background} provides background on network packets, self-supervised learning, and few-shot learning. Section~\ref{sec:data} describes the datasets and pre-processing steps. Section~\ref{sec:ThreatAssumptions} details the model detection assumptions. Section~\ref{sec:arch} details the proposed model architecture and training protocols. Section~\ref{sec:results} presents the evaluation and ablation results, and Section~\ref{sec:threat} discusses experimental results using AES and Fernet encryption. Section~\ref{sec:discussion} provides further discussion, and Section~\ref{sec:conclusion} concludes the paper.

\section{Related Work}
\label{sec:rw}
In this section, we discuss related works employing modern techniques for detecting and mitigating cyber threats, focusing on approaches that leverage machine learning, deep learning, attention mechanisms, and FSL to improve accuracy in various detection tasks, summarized in Table \ref{tab:comparative-summary}.

\subsection{Machine/Deep Learning Attack Detection}
Aceto et al. \cite{b3} delve into the application of deep learning in classifying mobile network traffic, highlighting the capability of deep learning systems to process network traffic independently of port information, thus distinguishing between different app-generated traffic effectively. In \cite{Patheja}, researchers explored various methodologies, including self-taught learning and softmax regression, utilizing the NSL-KDD dataset for benchmarking. The research applies supervised learning to analyze over 41 statistical attributes, excluding payload byte details. Doshi et al. \cite{DDOS} focus on utilizing machine learning to detect Distributed Denial of Service (DDoS) attacks, using a combination of four machine learning algorithms and a simple neural network to differentiate between normal and DDoS traffic in Internet of Things (IoT) environments. They emphasize the creation of a new dataset for algorithm testing, noting the significant size and imbalance between malicious and benign data samples.

In another study \cite{Network}, machine learning is employed for IP traffic classification within a 4G network context, with the researchers generating a dataset specific to 4G traffic and applying machine learning algorithms to its packet content. This research specifically excludes analysis of malicious traffic, focusing instead on general IP traffic flow. Similarly, Almaraz-Rivera et al. \cite{IOT} investigated the use of machine and deep learning methods for detecting DDoS attacks, relying on statistical data like packet counts, bytes, and the duration of transactions rather than raw packet bytes. 

Stein et al. \cite{CANRnn} demonstrated that CAN payloads in vehicles can be classified as benign or malicious using a supervised deep learning approach with a Recurrent Neural Network (RNN). However, it's important to note structural differences: CAN bus messages have an 8-byte payload limit, while IP network packets typically support up to 1500 bytes. Recent works include the DEEPShield system, an ensemble learning approach designed to detect solely DDoS attacks in IoT environments by integrating Convolutional Neural Networks (CNNs) and Long Short-Term Memory (LSTM) networks with a network traffic analysis module \cite{saiyed2024deep}. While these methods show the ability to generalize known attack types, it remains a critical consideration to detect novel and unseen types entering a network. Furthermore, these methods predominantly rely on statistical or high-level features of network traffic, overlooking the potential of raw packet-level byte data, which can provide more intricate representations of the network content. 

\subsection{Attention-based Attack Detection}
Researchers have recently proposed several innovative approaches for malware detection and classification. For example, ResNeXt+ \cite{transaction} integrates various plug-and-play attention mechanisms into the ResNeXt architecture to improve the identification of malware by converting malware PE/APK files into RGB information. The authors highlighted the model's effectiveness across multiple datasets and various scenarios. In another notable work, Zhou et al. \cite{Wf-transformer} proposed the use of Transformers in the cybersecurity domain, showcasing advancements in website fingerprinting over previously used CNNs by effectively capturing long-term dependencies in traffic traces through self-attention mechanisms. This approach has demonstrated promising results in both closed-world and open-world scenarios, providing enhanced performance in WF tasks. Similarly, Ravi and Alazab \cite{ravi2023attention} proposed an attention-based CNN approach for robust malware classification. The method enhances malware detection by integrating attention mechanisms into the CNN architecture, helping improve feature extraction and classification accuracy. Researchers in \cite{stein2024towards} employ a Large-Language Model (LLM) as a feature extracting backbone to extract low-dimensional embeddings of raw packet data for downstream classification tasks. Zhang et al. \cite{zhang2020ransomware}  addressed the challenge of detecting and classifying ransomware by proposing a static analysis framework based on a partitioning strategy combined with a self-attention-based CNN (SA-CNN) and a bi-directional self-attention network. While these attention-based approaches tend to improve performance over classic deep-learning techniques, a common limitation remains: the reliance on large labeled datasets. The problematic nature of accessing extensive labeled data, as well as the models' limited ability to adapt and classify previously unseen malware classes, continues to be an overlooked challenge in the field. 

\subsection{Few-Shot Learning Attack Detection}
FSL's aim is to learn from limited amounts of labeled data. This is particularly important for intrusion detection, where obtaining large, labeled datasets for new or emerging attack types is difficult. The cyber domain has began to implement FSL techniques to reduce the need for labeled datasets. For example, a FSL method was proposed to address the data imbalance issue in network intrusion detection by extracting statistical-based network flow features using a CNN and performing FSL analysis downstream \cite{yu2020intrusion}. Similarly, researchers implemented a Siamese CNN to extract features for anomaly detection in industrial cyber-physical systems \cite{zhou2020siamese}. The CNN features were once again used for FSL downstream to measure similarities between samples, enabling effective detection of anomalies with minimal labeled data. Lastly, another proposed FSL-based intrusion detection model tailored for IoT environments addressed the challenge of limited labeled data and emerging, zero-day attacks \cite{lu2023few}. 

Although these works produce promising results, they primarily rely on simple CNN-based feature extractors. Simple CNN-based extractors often struggle to capture long-term dependencies and subtle variations within packets, limiting their ability to generalize to unseen attack types effectively. SSL addresses these shortcomings by learning robust and generalizable feature representations based off attention. Additionally, prior FSL methods simulate unseen attacks by re-splitting known classes into different training and testing subsets. The model still has exposure to all of the classes distributions, just with limited examples from each. In contrast, our approach completely excludes one of the attack classes during SSL pre-training and then evaluates the model’s ability to detect the unseen class along with the seen classes in the FSL evaluation step. This approach is similar to our previously published work \cite{stein2024towards}, where one class is explicitly excluded from training to simulate the unseen scenarios. However, in our previous work we relied on \textbf{supervised} learning to pre-train the LLM on scratch from known malware types. Our work in this manuscript differs since we employ SSL techniques to pre-train the transformer on \textbf{unlabeled} data. This helps demonstrate that SSL can learn meaningful representations from the packet data without relying labeled data, a crucial step in real world applications. Overall, this approach ensures that the untrained class remains completely unseen during the training phase.

\section{Background}
\label{sec:background}

The proposed methodology implements multiple learning criterion throughout the study. In this section, we discuss the necessary preliminary information that sets the foundation for understanding our approach.

\subsection{Fundamentals of Network Packets}

\begin{figure}[t!]
  \centering
  \includegraphics[width=\columnwidth]{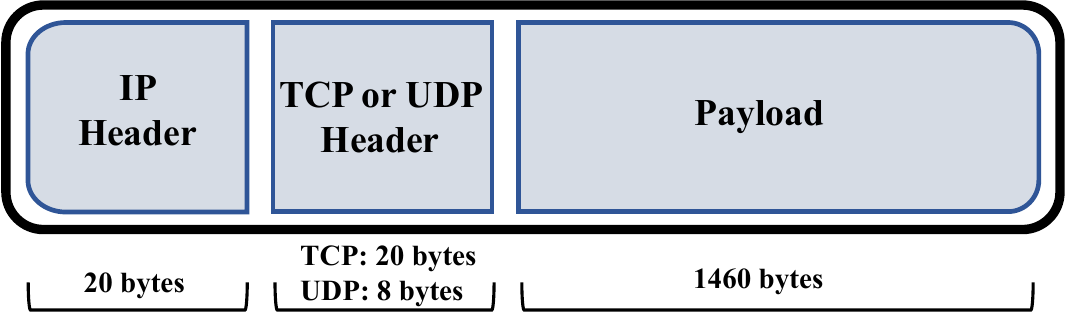} 
  \caption{Network packet structure.}
  \label{fig:stein1} 
\end{figure}

Network packets play a critical role in the transmission of data across networks. These packets contain information for routing and delivery, including the source and destination IP addresses. IP addresses identify the sending and receiving devices, and the source and destination ports, which specify the sending and receiving applications. These elements form what is known as the five-tuple: Source IP Address, Destination IP Address, Source Port, Destination Port, and Transport Protocol. The Transport Protocol component of the five-tuple specifies the type of protocol used for the connection. Transmission Control Protocol (TCP) and User Datagram Protocol (UDP), being the two most common transport layer protocols \cite{b9}, are the focus in this study. TCP is known for its reliability and ordered delivery of packets, establishing a connection-oriented session between communicating devices. TCP guarantees that data is received exactly as it is sent, making it ideal for applications where accuracy is necessary, such as web browsing and email transmission \cite{TCPvsUDP, kumar2012survey}. UDP, on the other hand, is a connectionless protocol known for its low-latency and efficient data transfer, without the same reliability guaranties as TCP. This is commonly used in streaming media, online gaming, and real-time communications where speed is more crucial than perfect delivery \cite{TCPvsUDP, kumar2012survey}. These two protocols account for the majority of network traffic at the transport layer. 

Given that the five-tuple function is to identify connections, it is important to note that packets also encompass network payloads. These payloads carry the substantive content being transmitted, including text, images, and files, from the source to the destination. Malicious actors often embed malware within these payloads, exploiting the data transfer process to compromise targeted systems \cite{b9, b10, IOT, farrukh2022payload}. In a given network packet, the maximum payload size is up to 1460 bytes. Payloads that exceed this limit are fragmented to ensure the complete transmission of data across the network, preventing any portion of the content from being lost during transit. Notably, the fragmentation of network payloads depends on the capacity of the network segment, and these fragments can be reassembled either at the endpoint or by intermediate devices. Furthermore, transformer-based architectures are well-suited for processing fragmented payloads since the self-attention mechanism allows the model to capture dependencies across separate fragments, therefore integrating information that is dispersed over multiple packets. Initially, raw packet and payload data exists in binary form and is then converted to hexadecimal format for various purposes. The hexadecimal representation offers a human-readable form, simplifying the process of data inspection and analysis for network experts. Fig. \ref{fig:stein1} shows a basic diagram of the header and payload bytes with a Maximum Transmission Unit (MTU) of 1500 bytes. The MTU represents the largest size of an Ethernet packet or frame that can be transmitted over a network. \emph{Furthermore, it is important to note that DPI operates under the principle that the majority of data transmissions occur without encryption, which enables direct inspection of packet content}\cite{deri2021using, dijk2021detection}.

\begin{figure}[t!]
  \centering
  \includegraphics[width=\columnwidth]{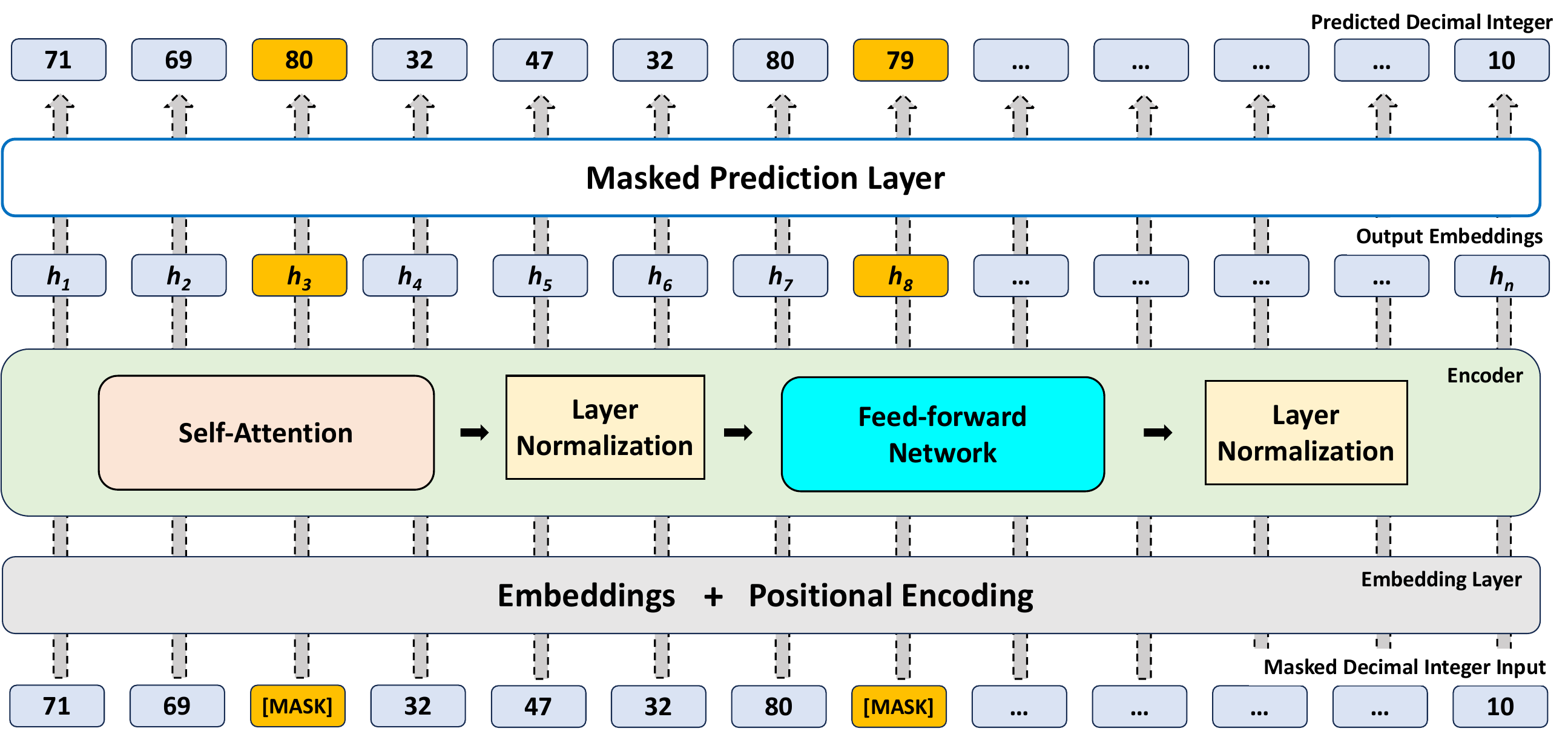} 
  \caption{Self-supervised learning with Masked Language Modeling.}
  \label{fig:MLM_Fig} 
\end{figure}

\subsection{Self-Supervised Learning}
SSL learns a representation of the data via a pretext task. The pretext task is defined depending on the application and the type of the dataset. For example, the pretext task in natural language processing (NLP), which involves learning patterns from a sequential dataset (i.e. sentences), is predicting masked words in a sentence. Through prediction of the masked words, the internal transformer architecture learns to extract distributed representations of entire sentences without the masked word, such that the representation carries enough information to predict what the missing word is. Operating on the principle that the structure of the data can provide insights in an unsupervised manner, it seeks to reduce dependence on labeled datasets and learn to recover whole or parts of the features of its original input \cite{SSL, SSL2}. Masked language modeling (MLM) is a key component in SSL and introduced a novel methodology to utilize unlabeled data \cite{b1}. MLM involves predicting a masked input within a sequence, allowing the model to consider context from both the left and right sides of any masked input. The bidirectional attention grants the model complete access to all surrounding tokens, enhancing the ability to understand the full context of the sequence \cite{b1}. We extend this concept to understand the relationship between the bytes in a packet sequence based on the surrounding context. This approach enhances the model's ability to generalize from the unlabeled data to unseen examples, improving its predictive accuracy on the raw packet bytes

As shown in Fig. \ref{fig:MLM_Fig}, the MLM process begins with masked input values, where specific positions (i.e. [MASK]) are hidden from the model. The input values are passed through an embedding layer combined with positional encoding to preserve the order sequence. The transformer encoder then processes the masked sequence through self-attention and feed-forward networks, followed by layer normalization. The resulting output embeddings are sent to a masked prediction layer, which predicts the missing values based on the learned contextual representations. The predicted values are compared with the actual masked values using a cost function, and the resulting loss is backpropagated through the model to update all parameters. This modeling approach is effective for tasks demanding a global understanding of an entire sequence's context.

\subsection{Few-Shot Learning}
Similar to how humans are able to recognize objects after observing them once, FSL is a specialized approach designed to enable models to learn and generalize from very limited amounts of labeled data \cite{OneShot, PrototypicalNetworks, LaplacianNetworks}. In FSL, a classifier needs to accommodate for new classes not seen in training, given only few examples from each of those classes. One-shot learning is a form of few-shot learning where the model needs to learn given only one example per class \cite{OneShot, stein2025transductive}. These algorithms must be able to extract rich contextual understanding from the limited amount of seen examples to generalize to new and unseen instances.

Snell et al. \cite{PrototypicalNetworks} introduced Prototypical Networks for FSL. The main idea behind this research was to show that data points belonging to the same class should be close in distance to each other in an embedding space. The authors suggested that each class should have a single representative or prototype, and this prototype is the mean of all data points in that class once they have been projected into a shared embedding space. Support and query sets are then used to simulate the few-shot learning scenario during training and testing. The support set consists of a small number of labeled examples for each class and are used to define the prototype by computing the mean vector of the embedded points for each class. The query set contains unlabeled examples that the model needs to classify. These examples are different from those in the support set and are used to evaluate the performance of the model. Once the query and support sets are embedded into the space, the model computes the Euclidean distance between each query point and the prototypes. The query points are then classified based on the nearest prototype in the space.

\section{Data}
\label{sec:data}

\begin{figure*}[t!]
\centering
\includegraphics[width=1.0\linewidth]{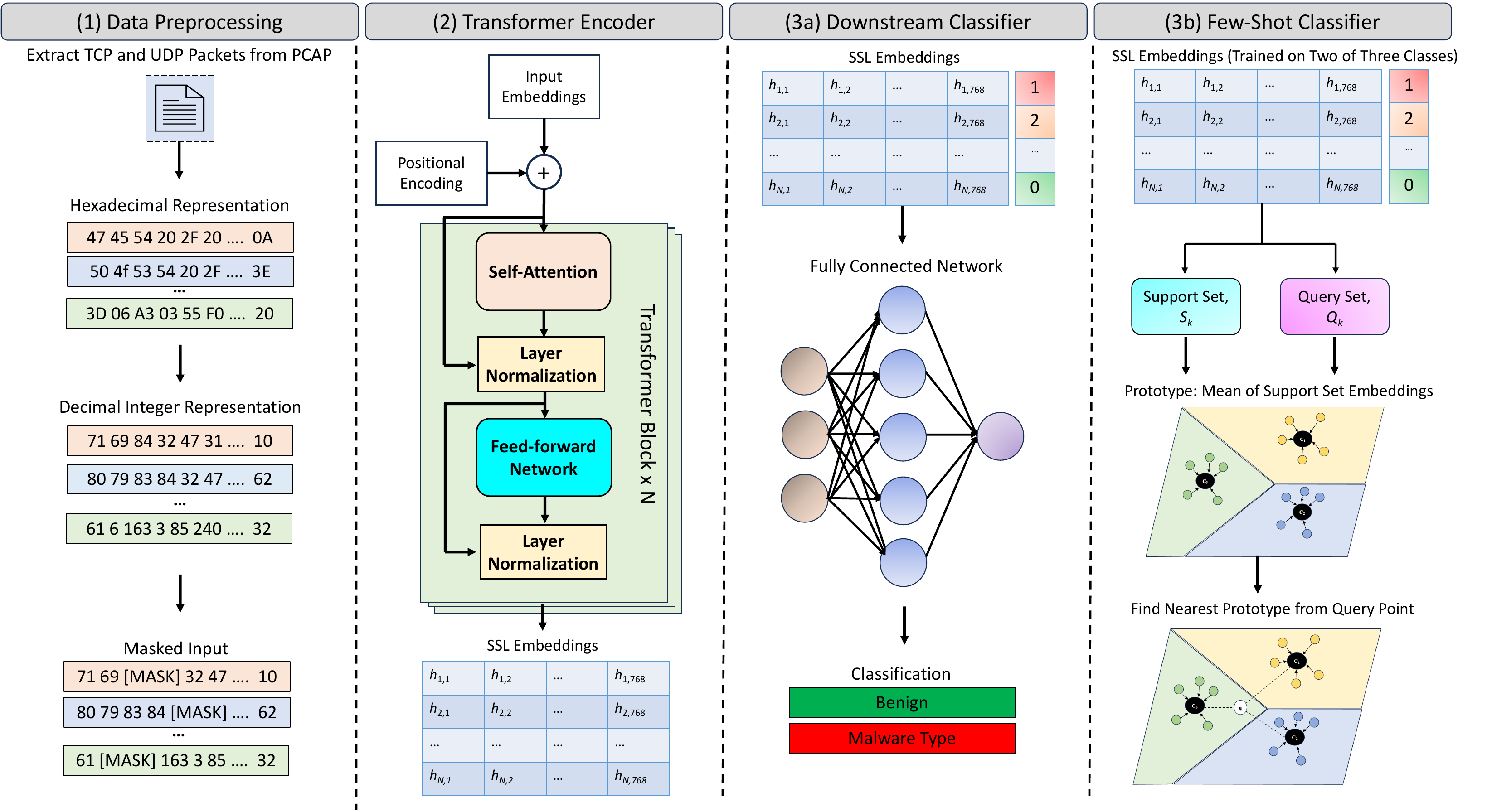}
\caption{The overall architecture of the proposed packet detection algorithm.}
\label{fig:Main_Fig}
\end{figure*}

In this section, we discuss the characteristics and pre-processing of two distinguished and well-recognized datasets: UNSW-NB15 \cite{UNSW} and CIC-IoT23 \cite{CICIOT}. These two datasets were chosen due to their comprehensive nature and relevance in the field of cybersecurity, particularly in the context of network intrusion detection and IoT security. Both datasets are available from their official websites, and the pre-processing steps outlined below allow for full reproducibility.

\subsection{UNSW-NB15}
The UNSW-NB15 dataset was developed to address the need for more extensive public intrusion detection network datasets. Although benchmark datasets such as the KDDCUP99\cite{KDD9} were prominently used for many years, the UNSW-NB15 tackles the shortcomings these older datasets have, such as redundancy and missing records. The UNSW-NB15 comprises 100 GB of raw network traffic captured in PCAP format, featuring a balanced mix of real and simulated modern attack scenarios. For practicality and to ensure efficiency throughout the study, from the available 100 GB of PCAP data, a subset of 4 GB was selected. This subset includes diverse traffic scenarios, representing a comprehensive coverage of network attacks and behaviors that are available in the data. 

The primary input to the model consists of packet bytes, represented as decimal integers. To ensure a balanced dataset and minimize bias, an equal number of malicious packets from three selected attack types are randomly sampled. This step is crucial to avoid overrepresentation of any single attack, which could otherwise lead to an increase in false positives or negatives during classification. Duplicate packets are also removed to prevent unintended biases. For multi-class experiments, we focus on three attack categories—Fuzzers, Exploits, and Generic—due to their payload complexity, distinguishing them from simpler, volume-driven attacks like Distributed Denial of Service (DDoS).

\subsection{CIC-IoT23}
The CIC-IoT23 dataset is a comprehensive resource designed to explore vulnerabilities in Internet of Things (IoT) security. It includes seven broad categories of attacks—DDoS, Brute Force, Spoofing, Denial of Service (DoS), Recon, Web-based attacks, and Mirai malware—further divided into over 30 specific IoT-focused attack types. The dataset represents detailed network activity captured from a diverse IoT environment with 105 unique devices, making it a rich source of PCAP files that log both benign and malicious activities. 

The pre-processing for CIC-IoT23 closely parallels that of UNSW-NB15. Transport layer payloads are extracted from TCP and UDP packets in randomly selected PCAP files. These extracted bytes are initially converted to hexadecimal format, with duplicates removed to maintain data integrity. The cleaned data is then converted into decimal integers, and each malicious class is assigned a unique label. To maintain balance in the dataset and prevent model bias, we randomly sample an equal number of malicious entries from each attack class. For this study, we narrow our focus to three specific attack types from CIC-IoT23: Backdoor Malware, Vulnerability Attacks, and Brute Force Attacks. These attacks rely on the content of the packet to execute their malicious objectives, making them comparable to the chosen attacks from the UNSW-NB15 dataset.

\section{Threat Detection Assumptions}
\label{sec:ThreatAssumptions}

Our model focuses on active, traffic-based detection by analyzing the content of network packet bytes. The primary goal is to identify malicious content before execution by inspecting packet byte sequences during transmission. To clarify the scope and limitations of this work, we explicitly define the assumptions associated with our approach.

We assume access to active network traffic, specifically focusing on packet content from TCP and UDP transmission protocols. A significant portion of data transmissions occurs in plaintext \cite{deri2021using, dijk2021detection}, which permits direct inspection of packet content. We also assume that analyzing the complete payload, even when it is fragmented across multiple packets, is sufficient for detection. With these assumptions, the proposed method offers several capabilities. By leveraging SSL during training, the model can learn robust representations of network bytes without labels, enabling the detection of malware content early in active traffic. Furthermore, the proposed method generalizes to unseen malware classes using FSL, where limited labeled examples are sufficient to classify introduced attack types. The reliance on packet data also allows for efficient, real-time processing, making the approach scalable to high-throughput network traffic. Lastly, since the model assumes plaintext packet bytes, its effectiveness may diminish for encrypted traffic. To mitigate these risks, we evaluate the impact of encryption on detection performance by applying two symmetric encryption methods and repeating the method.

\section{System Architecture and Model Training}
\label{sec:arch}
In this section, we describe the system architecture, training, and evaluation of our model for network traffic classification. Our method operates in multiple stages: we first pre-train a byte-level transformer encoder with SSL through a MLM objective \textit{from scratch} on unlabeled packet bytes, producing packet embeddings from the final hidden states of the transformer. Next, to validate the learned embeddings, we implement a lightweight downstream classifier on top of the frozen encoder for binary and multiclass malware detection and classification under the assumption that sufficient labeled data exists for conventional downstream classification. The overall architecture of the proposed method can be referred to in Fig. \ref{fig:Main_Fig}.

\subsection{Transformer Encoder}
Transformer-based models have been powerful in natural language processing tasks due to their ability to uncover the complex relationships between words in sequences of textual data. Although individual bytes in a network packet may not inherently carry semantic meanings like words in a sentence, their sequences, patterns, and relative positions can encapsulate information about the nature of the packet. Leveraging transformer's capability to capture contextual patterns, we show that it can be effectively applied to discern distinctions in packet sequences. This process is powered by the self-attention mechanism \cite{Attention}, enabling each byte in the input sequence to reference and weigh other bytes within that sequence while formulating its output representation. The mechanism determines a significance score for each byte in the sequence for the degree of attention it should receive. The transformer encoder architecture can be divided into these main parts:

\vspace{1mm}
\noindent \textbf{Embedding Layer:} The embedding layer is the initial conversion phase between the input data and the transformer. The input, $\mathbf{X} = [x_1, x_2, \ldots, x_N]$, is transformed into the token-embedding matrix $\mathbf{Z} \in \mathbb{R}^{N \times d_{\text{model}}}$, where $N$ is the number of decimal integer values in the sequence and $d_{\text{model}}$ is the dimension of the embedding vectors. The embedding layer converts these input sequences, encoded and padded during the preprocessing phase, into embedding vectors of fixed size 768. This conversion is performed by the learnable embedding matrix $\mathbf{W}_E$, which maps the fixed input tokens to their corresponding embedding vectors in the feature space. The output, \(\mathbf{Z}\), is given by:

\begin{equation}
\mathbf{Z} = \mathbf{W}_E\mathbf{X},
\end{equation}

\noindent where \( \mathbf{W}_E \) is the embedding matrix, with each distinct \( \mathbf{W}_E \) corresponding to a particular instantiation of the embedding parameters that will be learned during training.

\vspace{1mm}
\noindent \textbf{Positional Encodings}: 
For transformers to understand and preserve the order of sequences, positional encodings are added to the input embeddings of the model. These encodings provide a unique position to each element in the sequence, allowing the model to recognize the position of each byte in the network packet and their relation to others within the same sequence. This aids the self-attention mechanism by treating each input independently and would otherwise not maintain information about sequence order. 

The positional encoding vector is designed to have the same dimension, \(d_{\text{model}}\), as the embedding layer, allowing the two to be summed together. The encoding for position \(pos\) and dimension \(i\) is calculated using sine and cosine functions:

\begin{align}
PE_{pos,2i} &= \sin\left(\frac{pos}{10000^{2i/d_{\text{model}}}}\right) \\
PE_{pos,2i+1} &= \cos\left(\frac{pos}{10000^{2i/d_{\text{model}}}}\right)
\end{align}

This ability lets the model shift its attention globally across the sequence, paying more attention to certain parts based on what they are and their position. This makes the model more efficient at recognizing patterns within the sequence.

\vspace{1mm}
\noindent \textbf{Multi-Head Self-Attention Mechanism}:
The self-attention mechanism is what sets transformers apart from other traditional deep learning architectures. The self-attention mechanism allows the model to capture dependencies and relationships across the entire input sequence, regardless of their distance from each other. The ability to weigh the importance of different bytes based on both their content and their positions within the sequence is what makes the self-attention mechanism powerful for the given sequential task. This is achieved through the following steps:

\subsubsection{Input Linear Transformation}
Once the input sequence has been converted into the embedded matrix \( \mathbf{Z} \), each vector \( \mathbf{z}_i \) within this matrix is then transformed into three distinct vectors: queries (\( \mathbf{Q} \)), keys (\( \mathbf{K} \)), and values (\( \mathbf{V} \)), using the respective weight matrices \( \mathbf{W}_Q \), \( \mathbf{W}_K \), and \( \mathbf{W}_V \), which are learned during the training process. These transformations can be described by the following equations: 
\begin{align}
\mathbf{Q} &= \mathbf{W}_Q\mathbf{Z} \\
\mathbf{K} &= \mathbf{W}_K\mathbf{Z} \\
\mathbf{V} &= \mathbf{W}_V\mathbf{Z}
\end{align}
where \( \mathbf{Z} \) is the matrix of embedded vectors, and \( \mathbf{W}_Q \), \( \mathbf{W}_K \), and \( \mathbf{W}_V \) are the weight matrices that transform the embedded vectors into queries, keys, and values, respectively.

\subsubsection{Computation of Attention Scores}
Attention scores are computed by taking the dot product of the query with all keys, scaled by \( \frac{1}{\sqrt{d_k}} \), where \( d_k \) is the dimension of the key vectors. The softmax function is applied to these scores to obtain a probability distribution.
\begin{align}
\text{Attention}(\mathbf{Q}, \mathbf{K}, \mathbf{V}) &= \text{softmax}\left(\frac{\mathbf{Q}\mathbf{K}^T}{\sqrt{d_k}}\right)\mathbf{V}
\end{align}

The final output represents an aggregation of the input features, detailing the most relevant information for each byte in the sequence. This mechanism allows the model to capture both the global understanding and complex information necessary for understanding the packet input patterns.

\subsubsection{Masked Prediction Layer} 
The primary function of the masked prediction layer is to predict the masked bytes in the input sequence. This layer converts the encoded embeddings into logits corresponding to the vocabulary size. These logits are then passed through a softmax function to produce a probability distribution over the entire vocabulary for each masked position. These probabilities represent the model's predictions for the masked bytes in the input sequence.

\subsection{Model Training}

\begin{algorithm}[t!]
\caption{\textbf{Self-Supervised Training and Embedding Extraction}}
\label{alg:ssl_training}
\begin{algorithmic}[1]

\STATE \textbf{Input:} Set of network packets \( P = \{p_1, p_2, \ldots, p_n\} \)
\STATE \textbf{Output:} Packet embeddings \(\mathbf{E}\) for downstream tasks
\STATE \textbf{Step 1: Packet Byte Masking}
\FOR{each packet \(p_i \in P\)}
    \STATE Convert \(p_i\) to decimal integers \(\mathbf{x} = (x_1, x_2, \ldots, x_l)\), where \(x_j \in [0,255]\).
    \STATE Let $\alpha = 0.15$. Randomly select $\lceil \alpha \cdot l\rceil$ positions from $\{1,2,\ldots,l\}$ to form the mask set $R$.
    \FOR{each index \(j \in R\)}
        \STATE With 80\% probability: \(x_j \gets [\text{MASK}]\)
        \STATE With 10\% probability: \(x_j \gets \text{Random token} \in [0,255]\)
        \STATE With 10\% probability: \(x_j \gets \text{Unchanged}\)
    \ENDFOR
    \STATE Pad  \(\mathbf{x}\) to fixed length \(L_{\max}\) to form \(p_i'\).
\ENDFOR

\STATE Let \(P' = \{p_1', p_2', \ldots, p_n'\}\) be the masked dataset.

\STATE \textbf{Step 2: Transformer Model Training}
\STATE For each \(p_i' \in P'\), embed via \(\mathbf{W}_E\) to get \(\mathbf{Z}_i\). Add positional encodings \(\mathbf{PE}\). 
\STATE Train the transformer using self-attention to predict masked bytes:
\[
\min_{\theta} \ \mathcal{L}(\theta),
\]
where \(\mathcal{L}\) is the cross-entropy over masked positions. Use Adam for optimization.


\STATE \textbf{Step 3: Embedding Extraction}
\STATE After convergence freeze transformer's parameters $\theta$ \vspace{-1pt}.
\STATE For any input packet $p_i$, feed forward to obtain last hidden state embeddings $\mathbf{H}_i$.
\STATE Compute the packet embedding $\mathbf{e}_i$ by mean pooling the rows of $\mathbf{H}_i$.
\STATE \textbf{Return} the set of packet embeddings $\{\mathbf{e}_i\}$.

\end{algorithmic}
\end{algorithm}

\begin{algorithm}[t!]
\caption{\textbf{Few-Shot Learning for Malware Classification}}
\label{alg:few_shot}
\begin{algorithmic}[1]

\STATE \textbf{Input:}
\begin{itemize}
    \item Support set $S$ containing examples from three malware classes (one of which is unseen/untrained).
    \item Query set $Q$ (unlabeled samples for evaluation).
\end{itemize}

\STATE \textbf{Output:} Predicted labels $\hat{y}_q$ for each query point $q \in \mathcal{Q}$.

\STATE \textbf{Step 1: Episodic Training}
\FOR{training episode = 1 \textbf{to} max\_episodes}
    \STATE Randomly generate a support set $S_k$ and query set $Q_k$ for each class $k = 1,\dots,K$.
    \FOR{each class $k \in \{1,\ldots,K\}$}
        \STATE Compute the prototype:
        \[
            c_k = \frac{1}{|\mathcal{S}_k|} \sum_{(x_i, y_i)\,\in\,\mathcal{S}_k} f_\theta(x_i).
        \]
    \ENDFOR
    \FOR{each query sample $(x_j, y_j) \in Q_k$}
        \STATE Compute embedding: $\mathbf{e}_j = f_\theta(x_j)$
        \STATE Assign predicted label: $\hat{y}_j = \arg\min_{k}\, d\bigl(\mathbf{e}_j, c_k\bigr)$
    \ENDFOR
    \STATE Compute the loss $J$ over the query predictions.
\ENDFOR

\STATE \textbf{Step 2: Few-Shot Inference}
\STATE Compute prototypes $c_k$ for each class $k$ using the support set $S$.
\FOR{each query sample $q \in Q$}
    \STATE Compute embedding: $\mathbf{e}_q = f_\theta(q)$
    \STATE Assign predicted label: $\hat{y}_q = \arg\min_{k}\, d\bigl(\mathbf{e}_q, c_k\bigr)$
\ENDFOR

\end{algorithmic}
\end{algorithm}

The self-supervised transformer model is trained on packet data without labels, leveraging a MLM objective at the byte-level. Following standard masking strategies \cite{b1}, 15\% of bytes in each input sequence are randomly selected for masking. Out of those selected for masking, 80\% are replaced with a special mask token, 10\% are replaced by a random byte sampled from the vocabulary, and the remaining 10\% are left unchanged. Although some selected bytes are not physically replaced by the mask token, they are still considered “masked” from the model’s perspective. Therefore, the model learns to predict the original values of these masked bytes using the context provided by the unmasked bytes. Cross-entropy loss is employed as the primary cost function for masked byte prediction. The cross-entropy can be defined as:


\begin{equation}
\mathcal{L} = -\frac{1}{N_{\text{masked}}} \sum_{i=1}^{N} \sum_{c=1}^{V} y_{i,c} \log(p_{i,c}), 
\end{equation}

\noindent where $V$ is the number of unique packet elements, $y_{i,c}$ represents the ground truth for the $i$-th masked element, and $p_{i,c}$ is the predicted probability for each element $c$ at position $i$. $N$ is the total count of masked elements across the batch, and $N_{\text{masked}}$ is used to normalize the loss. This approach focuses the model on predicting masked elements using the context provided by the unmasked elements, leading to learn the representation of the data’s underlying structures.

The Adam optimizer is used with a learning rate of $2e-5$ and provides weight decay regularization, which prevents the weights from growing too large to help mitigate overfitting \cite{kingma2017adammethodstochasticoptimization}. Overfitting occurs when a model fits too closely to the training dataset, leading to poor performance on the testing data. A scheduler for learning rate decay is also used to reduce the learning rate over the training period. The scheduler gradually increases the learning rate from zero to a specified learning rate during the warmup period, then linearly decreases the learning rate over the remaining epochs of training \cite{huggingface_optimizer}. The experiments were conducted on NVIDIA GeForce RTX 2080 GPUs. This hardware enabled us to use significant computational power, facilitating quicker processing and learning. 

\subsection{Embedding Extraction}
After the training is complete, we process each input packet sequence through the trained transformer's final hidden layer to produce a contextualized embedding vector for every byte in the input sequence. To obtain a vector representation for each packet, we perform mean pooling over the sequence dimension of the last hidden states. Let \( \mathbf{H} \in \mathbb{R}^{N \times d_{\text{model}}} \) represent the last hidden state, where \( N \) is the sequence length and \( d_{\text{model}} \) is the embedding dimension. We compute the final packet embedding \( \mathbf{e} \in \mathbb{R}^{d_{\text{model}}} \) as:

\begin{equation}
\mathbf{e} = \frac{1}{N}\sum_{i=1}^{N} \mathbf{H}_i.
\end{equation}

This operation aggregates the information across all bytes in the sequence, resulting in a single embedding vector that captures the semantic features of each packet. These embeddings are used as input for downstream tasks, such as the fully-connected classifiers. For the FSL task, when a novel class is introduced that is excluded from SSL pre-training, its packet sequences are still passed through this same final hidden layer to extract embeddings. This enables the transformer’s learned feature space to generalize to novel or unseen classes. Algorithm \ref{alg:ssl_training} details the masking, training, and extraction of the SSL model. 

\subsection{Model Configuration}

The selected hyperparameters of the model include the number of unique bytes (\textit{i.e.} 258), hidden size, number of hidden layers, number of attention heads, the intermediate size, maximum length of the input payload,  and the number of labels for this application, and are presented in Table~\ref{tab:bert_config}.

The unique bytes parameter represents the total number of unique elements found in the input data. Here, the total is 258, accounting for each unique hexadecimal value and additional pad, unknown, and masking integers, following standard self-supervised practices. The hidden size parameter is essential in determining the dimension of the hidden state. The number of hidden layers represents the number of stacked transformer blocks present in the model. Complementing these blocks is the number of attention heads, which control the model's ability to attend to different segments of the input. This is pivotal for the model to comprehend and learn from the input data effectively. The intermediate size determines the dimension of the intermediate later in the network within each transformer layer. This value helps ensure that the network can process and transform the input data at each layer effectively. The maximum position embeddings denote the maximum length of the input sequence. For this study, we do not restrict the maximum packet length, which is set to 1500. It is important to note that transformer architectures are inherently capable of processing sequences that extend beyond a single packet's MTU, meaning that if a payload exceeds 1460 bytes and is split across multiple packets, the transformer can process it as a continuous stream of bytes across packet boundaries. However, increasing the maximum input length significantly raises computational resource requirements, such as memory usage and processing time. These configurations remained consistent through our study of extracting the packet byte embeddings.

\begin{table}[t!]
\centering
\caption{Model Configurations}
\label{tab:bert_config}
\begin{tabular}{|c|c|}
\hline
\textbf{Configuration Parameter} & \textbf{Value} \\
\hline
unique\_bytes           & 258  \\
hidden\_size            & 768  \\
num\_hidden\_layers     & 12   \\
num\_attention\_heads   & 12   \\
intermediate\_size      & 3072 \\
max\_position\_embeddings & 1500 \\
\hline
\end{tabular}
\end{table}

\subsection{Downstream Task}
After the extraction of the embeddings from the transformer trained with SSL approaches, a downstream task is applied for the purpose of malware classification to test the quality of the embeddings. In this paper, we assume \emph{two downstream task settings:} 1) sufficient annotated malware samples are available, enabling the training of a conventional classifier; and 2) only a few annotated samples are available, leading to the implementation of FSL.

In the first setting, we assume that a sufficient amount of labeled data is available. This is tasked with interpreting and classifying the embeddings to distinguish between benign and malicious network traffic and classifying the type of malwares. To achieve this, a \textbf{mutli-layer (M.L.)} classifier architecture is employed on top of the transformer model. 






\begin{table}[t!]
  \centering
  \caption{Downstream M.L. classifier architecture}
  \label{tab:downstream_arch}
  \begin{tabular}{lll}
    \toprule
    Layer                         & Output dim. & Components                   \\
    \midrule
    FC: $d_{\text{model}}\to512$  & 512         & ReLU, BatchNorm, Dropout(0.5) \\
    FC: $512\to128$               & 128         & ReLU, BatchNorm              \\
    FC: $128\to C$                & $C$         & softmax                      \\
    \bottomrule
  \end{tabular}
\end{table}

In this architecture, shwon in Table \ref{tab:downstream_arch}, the input embedding vector of dimension \(d_{\text{model}}\) is first passed through a fully connected linear layer that projects it into a 512-dimensional space. This is followed by batch normalization, a ReLU activation to introduce non-linearity, and a dropout layer with a dropout rate of 0.5 to prevent overfitting. The network then reduces the 512-dimensional output to 128 dimensions through a second linear layer, again applying batch normalization and a ReLU activation to maintain stability and non-linearity in the learning process. Finally, a linear layer maps the 128-dimensional vector to \(C\) output neurons, where \(C\) corresponds to the number of classes. A softmax activation function is applied to the final output to generate a probability distribution over the classes.

The model is trained using cross-entropy loss and the Adam \cite{kingma2014adam} optimizer is implemented with a learning rate of $0.01$. A learning rate scheduler is also incorporated into training, ensuring the model adapts its learning process over time. After training, the classifier interprets the embeddings to distinguish between benign and malicious packets. The classifier outputs a probability distribution across all defined categories, such as benign or malicious for binary, and types of malware for multiclass.


\section{Few-Shot Learning Architecture}


Our framework begins by pre-training the transformer model using SSL to extract the robust embeddings from network traffic data. When sufficient labeled data exists, these embeddings are used to train a downstream classifier for malware detection and classification. This classifier is highly effective for known malware types when there exists a large number of annotated data samples. \ul{However, when a new malware variant occurs in a network, often only a few labeled examples are available for these novel threats. In such cases, it is impractical to train a new classifier from scratch due to the scarcity of data.} To address this challenge, we employ a FSL approach that leverages robust representations learned during pre-training on known malware types. These representations enable the adaptation to novel malware classes with only a few annotated examples by transferring prior knowledge to classify new threats.

In the FSL scenario (\ul{the second setting}), once the embeddings of the packets are extracted using the pre-trained transformer, the support and query sets are randomly selected. Support sets consists of a limited number of labeled examples from each class, while the query set contains unlabeled examples that the model attempts to classify based on the knowledge learned from the support set. The model uses a limited number of support examples (shots) from each class (ways) to form class-specific prototypes. These prototypes represent the central tendencies of each class in the embedding space from the self-supervised features. The few-shot framework employs a fully connected linear layer, tasked with transforming the input embeddings to a representation space. The model is also trained using episodic training. Episodic training is a method used in few-shot learning to prepare models by organizing training into a series of learning problems, known as episodes, to mimic the circumstances encountered during testing \cite{episodic}.

In each episode, we randomly select \( K \) classes from the dataset. For each class \( k \), two distinct sets are sampled: a support set \( S_k \) comprising \( N_S \) examples, and a query set \( Q_k \) comprising \( N_Q \) examples. The algorithm computes the class prototypes \( \mathbf{c}_k \) by taking the mean of the transformed feature vectors of the support set embeddings, \( S_k \). Once the class prototypes are established, the algorithm evaluates each query sample \( (x, y) \) in \( Q_k \), calculating the Euclidean distance from \( x \) to each class prototype \( \mathbf{c}_k \), and then assigns the class label of the nearest prototype to \( x \) . The update to the model parameters is based on the loss \( J \), which is formulated as the negative log-likelihood of the softmax probabilities over the distances between the query samples and the class prototypes. Specifically, for each query \( (x, y) \) in \( Q_k \), the loss \( J \) is updated according to the equation:
\begin{equation}
\scriptsize
J \leftarrow J + \frac{1}{K N_Q} \left[ d(f_\theta(x), c_k) + \log \sum_{k'} \exp(-d(f_\theta(x), c_{k'})) \right].
\end{equation}
This process is repeated across numerous episodes, allowing the model to generalize better to new configurations of class examples. The overall procedure of the proposed framework is highlighted in Algorithm 2. 

\begin{table*}[t!]
\centering
\caption{%
  Results of Binary and Multiclass Classification of Malware Types on the Test Dataset. 
  \newline
  \textit{Note:} \textbf{Bold} font indicates the highest performance, while \underline{underlined} denotes the second-best result in each column.
}
\label{tab:combined_binary}
\resizebox{\textwidth}{!}{%
\begin{tabular}{@{}lcccccccccccccccc@{}}
\toprule
\multirow{3}{*}{\textbf{Method}} & \multicolumn{8}{c}{\textbf{Binary Classification}} & \multicolumn{8}{c}{\textbf{Multiclass Classification}} \\
\cmidrule(lr){2-9} \cmidrule(lr){10-17}
& \multicolumn{4}{c}{\textbf{UNSW-NB15}} & \multicolumn{4}{c}{\textbf{CIC-IoT23}} & \multicolumn{4}{c}{\textbf{UNSW-NB15}} & \multicolumn{4}{c}{\textbf{CIC-IoT23}} \\
\cmidrule(lr){2-5} \cmidrule(lr){6-9} \cmidrule(lr){10-13} \cmidrule(lr){14-17}
& Acc. & Prec. & Rec. & F1. & Acc. & Prec. & Rec. & F1. & Acc. & Prec. & Rec. & F1. & Acc. & Prec. & Rec. & F1. \\
\midrule
1D-CNN \cite{b9} & \textbf{95.19} & \underline{94.89} & \underline{95.77} & \textbf{96.31} & \underline{82.81} & \underline{82.33} & \underline{82.51} & \underline{82.92} & 89.88 & 90.15 & 89.88 & 89.88 & \underline{76.18} & 77.99 & \underline{76.19} & \underline{76.37} \\
2D-CNN \cite{b10} & 92.95 & 92.91 & 92.91 & 92.91 & 78.89 & 79.26 & 78.89 & 78.80 & 87.73 & 88.55 & 87.73 & 87.75 & 69.81 & 76.48 & 69.81 & 70.02 \\
LSTM \cite{ali2022effective} & 91.47 & 90.41 & 93.25 & 91.81 & 79.22 & 78.96 & 80.31 & 79.63 & 83.26 & 83.98 & 83.26 & 83.42 & 61.47 & 61.97 & 60.89 & 61.45 \\
ResNeXt+ \cite{transaction} & 91.66 & 92.11 & 91.16 & 91.13 & 82.10 & 82.11 & 82.10 & 82.09 & \underline{90.05} & \underline{90.18} & \underline{90.05} & \underline{90.12} & 75.14 & \underline{79.14} & 75.14 & 75.29 \\
DEEPShield \cite{saiyed2024deep} & 92.77 & 89.99 & \textbf{96.66} & 93.20 & 75.56 & 73.23 & 81.44 & 77.12 & 84.84 & 85.78 & 84.84 & 84.79 & 69.41 & 73.69 & 69.41 & 69.52 \\
\midrule
\textbf{Proposed (M.L.)} & \underline{94.88} & \textbf{95.04} & 94.88 & \underline{94.88} & \textbf{83.25} & \textbf{83.25} & \textbf{83.25} & \textbf{83.24} & \textbf{91.29} & \textbf{91.51} & \textbf{91.29} & \textbf{91.29} & \textbf{77.88} & \textbf{79.55} & \textbf{77.88} & \textbf{78.03} \\
\bottomrule
\end{tabular}%
}
\end{table*}

\section{Results}
\label{sec:results}
In this section, we delve into the metrics to report performance, the state-of-the-art (SOA) benchmarks used for comparison, and the results of our proposed method. We pre-train the encoder once per experimental setting and then run each downstream classifier three times with different random seeds, reporting the mean test accuracy (and episode-averaged mean for FSL).

\subsection{Metrics}
To evaluate the performance of the proposed method, several metrics are taken into consideration. For this study, we consider accuracy, precision, recall, and F1-Score on the test dataset. True positives (TP) are instances when the model correctly identifies a packet as malicious, a true negative (TN) are instances where the model correctly identifies the packet as benign, false positives (FP) are instances where the model incorrectly identifies a benign packet as malicious, and false negatives (FN) when the model mistakenly identifies a malicious packet as benign. The accuracy is interpreted as the proportion of packets that the model correctly identifies, as either malicious or benign, out of all sampled packets, shown by \( {Accuracy} = {{TP + TN} \over {TP + TN + FP + FN}} \). Precision measures the proportion of packets identified by the model as malicious that are truly malicious, \( {Precision} = \frac{TP}{TP + FP} \). Recall evaluates the proportion of all malicious packets present that the model successfully identifies, \( {Recall} = \frac{TP}{TP + FN} \). The F1-Score is the harmonic mean of precision and recall, \( {F1} = 2 \cdot \frac{{Precision} \cdot {Recall}}{{Precision} + {Recall}} \). These metrics provide a comprehensive evaluation of our model's ability in accurately detecting and classifying malware. Lastly, for clarification, the test dataset used in our evaluation process differs from the training dataset. The test dataset consists of randomly selected, unseen samples and ensures that the model is using a distinct dataset to evaluate the performance.

\subsection{Benchmarks}
\begin{table}[t!]
\centering
\caption{Average Inference Time Per Sample (ms)}
\label{tab:inference_time}
\setlength{\tabcolsep}{5pt} 
\renewcommand{\arraystretch}{1} 
\begin{tabular}{@{}lcccc@{}}
\toprule
\multirow{2}{*}{\textbf{Method}} & \multicolumn{2}{c}{\textbf{UNSW-NB15}} & \multicolumn{2}{c}{\textbf{CIC-IOT23}} \\ 
\cmidrule(lr){2-3} \cmidrule(lr){4-5}
& \textbf{Binary} & \textbf{Multiclass} & \textbf{Binary} & \textbf{Multiclass} \\ 
\midrule
1D-CNN \cite{b9}            & 0.51 & 0.16 & 0.13 & 0.14 \\ 
2D-CNN \cite{b10}           & 0.61 & 0.64 & 0.60 & 0.59 \\ 
LSTM \cite{ali2022effective}             & 6.10 & 6.33 & 5.80 & 5.63 \\ 
ResNeXt+ \cite{transaction} & 1.98 & 1.71 & 1.93 & 1.95 \\ 
DEEPShield \cite{saiyed2024deep} & 0.34 & 0.40 & 0.28 & 0.29 \\ 
\midrule
\textbf{Proposed (M.L.)}     & \textbf{0.03} & \textbf{0.06} & \textbf{0.04} & \textbf{0.04} \\ 
\bottomrule
\end{tabular}
\end{table}

Central to the evaluation of our study is the role of the downstream classifier, which confirm the effectiveness of the transformer embeddings produced through SSL methods. To display our proposed method's robustness, we compare with other SOA learning models for traffic classification that are comparable to our study. The 1D-CNN model \cite{b9} employs one-dimensional convolutional layers to extract hierarchical patterns from raw network packet data, effectively capturing local dependencies within sequential data. The 2D-CNN model \cite{b10} extends this concept by mapping packet data into a two-dimensional grid structure, allowing it to capture spatial relationships and patterns that may not be evident in a sequential representation. The LSTM \cite{ali2022effective} aids in modeling long-term dependencies in sequential data, making it proficient for capturing temporal patterns in packet streams. ResNeXt+ \cite{transaction} builds on the ResNeXt architecture by incorporating an attention mechanism into the residual blocks for feature exaction. Lastly, DEEPShield \cite{saiyed2024deep} takes a hybrid approach, developing an ensemble method technique that includes a CNN and LSTM. All these models are evaluated using the same SSL-generated embeddings, ensuring a consistent feature representation across comparisons and showcasing the adaptability of our embeddings to diverse architectures.

\begin{table*}[t!]
\centering
\caption{Results of Few-Shot Classification of Malware Types on the Query Set}
\label{tab:few-shot}
\setlength{\tabcolsep}{6pt}
\renewcommand{\arraystretch}{1.2}
\begin{tabular}{@{}lllcclccl@{}}
\toprule
\multirow{2}{*}{\textbf{Dataset}} & \multirow{2}{*}{\textbf{Trained Classes}} & \multirow{2}{*}{\textbf{Unseen Class}} 
& \multicolumn{3}{c}{\textbf{1-shot}} & \multicolumn{3}{c}{\textbf{5-shot}} \\
\cmidrule(lr){4-6} \cmidrule(lr){7-9}
& & & \textbf{Acc.} & \textbf{F1} & \textbf{Unseen Acc.} & \textbf{Acc.} & \textbf{F1} & \textbf{Unseen Acc.} \\
\midrule
\multirow{3}{*}{\textbf{UNSW-NB15}}
& Exploits, Fuzzers        & Generic        & 83.04 & 82.39 & 82.82 & 89.12 & 89.10 & 89.95 \\
& Fuzzers, Generic         & Exploits       & 82.86 & 82.26 & 82.54 & 89.48 & 89.74 & 89.68 \\
& Exploits, Generic        & Fuzzers        & 79.42 & 78.59 & 77.45 & 87.89 & 87.65 & 87.16 \\
\midrule
\multirow{3}{*}{\textbf{CIC-IoT23}}
& Backdoor, Vulnerability  & Brute Force    & 58.23  & 57.19   & 56.08    & 71.37    & 71.08    & 70.67    \\
& Vulnerability, Brute Force & Backdoor     & 61.32  & 60.25    & 61.35    & 74.26 & 73.89   & 74.07    \\
& Backdoor, Brute Force    & Vulnerability  & 62.35   & 61.38   & 61.73    & 75.01    & 74.91    & 75.05   \\
\bottomrule
\end{tabular}
\vspace{2mm}
\end{table*}

\begin{figure*}[t!]
  \centering
  \includegraphics[width=1.0\textwidth]{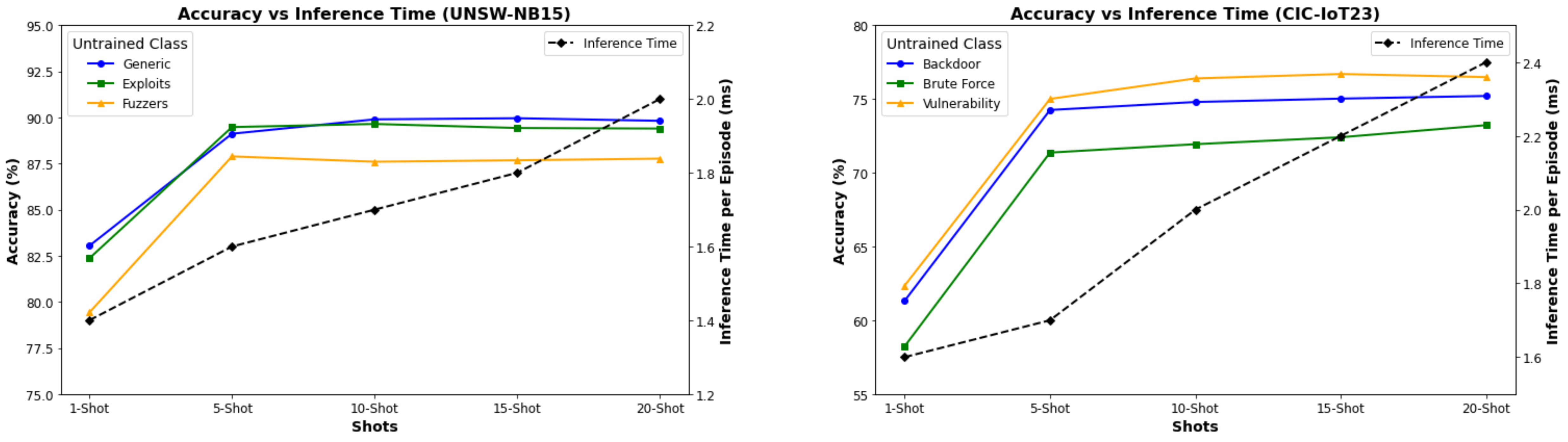} 
  \caption{Comparison of Accuracy and Inference Time UNSW-NB15 and CIC-IoT23 Datasets: Accuracy trends for different untrained classes are displayed alongside average inference time per episode for each number of shots.}
  \label{fig:AccInf} 
\end{figure*}

\subsection{Binary and Multi-Class Classification}
The data for the downstream classification task involving the SSL embeddings are divided into training, testing, and validation sets, with proportions of 70\%, 20\%, and 10\%, respectively. The performance of our proposed method was evaluated over 25 epochs, demonstrating improvements in accuracy, precision, recall, and F1-Score when using packet bytes as input, shown in Table \ref{tab:combined_binary}. 

For the binary classification task, the M.L. fully connected network demonstrates strong performance, with the M.L. classifier achieving SOA results. For the UNSW-NB15 dataset, the M.L classifier attains an accuracy of 94.76\%, precision of 94.95\%, and an F1-score of 94.81\%. Similarly, on the CIC-IOT23 dataset, the M.L classifier outperforms all other method's metrics, achieving an accuracy of 83.25\% and consistently high precision and recall scores. Similarly, when extending to a multiclass classification task to distinguish different malware types, a similar trend occurs. For example, the M.L. model achieves 91.51\% accuracy classifying Fuzzers, Exploits, and Generic attacks, and 78.03\% F1-score on CIC-IoT23 for classifying Backdoor, Vulnerability, and Brute Force attacks. Although the simpler M.L. model exhibits SOA performance, we also observe that other downstream architectures (i.e. CNN or DEEPShield) exhibit strong performance with the produced SSL embeddings. These findings confirm that the SSL transformer’s embeddings preserve valuable information about packet data, enabling robust performance across a range of classifier architectures.

The inference times per sample, shown in Table \ref{tab:inference_time}, further emphasize the efficiency of the proposed method. The M.L. fully connected networks has a significantly lower inference time compared to other benchmark models. The M.L network achieves an inference time of 0.03 ms and 0.04 ms for UNSW-NB15 and CIC-IOT23, respectively. In contrast, models such as LSTM and CNNs exhibit considerably higher inference times, with the LSTM model requiring over 6 ms per sample on UNSW-NB15. The faster inference time of the M.L. network make it ideal for real-time applications.

The enhancement in metrics can be attributed to several key factors inherent to our model's design and training approach. The extraction of the SSL embeddings allows our model to capture a deeper understanding of the dataset's intrinsic structure. Traditional supervised learning rely on labeled data, while SSL focuses on learning useful representations from unlabeled data via a pretext task, enabling the model to better generalize for the training and testing of the data. The improvement in metrics is also due to the transformer-based approach, which handles sequential data more efficiently due to the self-attention mechanism. This approach allows the model to evaluate each byte based on position within the entire sequence, giving it a unique advantage in identifying the significance of each byte relative to the whole sequence. The ability to globally understand the sequence helps identify when certain byte sequences or patterns might be indicative of malicious activity when appearing in specific contexts or configurations within the packet.



\subsection{Few-Shot Classification} 
\subsubsection{Experimental Setup}
The primary motivation for these experiments is to assess the model's ability to generalize to novel classes with limited supervision, a critical requirement for real-world scenarios where labeled data for all classes may not always be available. FSL allows us to test how well the SSL embeddings can adapt to previously unseen classes with very few labeled examples. To simulate this scenario, we train the SSL model exclusively on \textbf{two of the three malware classes} during pre-training, ensuring no prior exposure to the unseen class. 

We frame the problem as a 3-way classification task, testing both 1-shot and 5-shot settings, and using 15 query samples per episode. During each training episode, the model is presented with a small support set (either 1 or 5 labeled examples per class) and a set of query samples drawn from all three classes, including both previously seen (trained) and novel (untrained) classes. By exposing the model to queries from both known and new classes within the same episode, we ensure that it develops a global understanding of the entire class distribution, rather than focusing solely on the introduced class. Furthermore, we report the accuracy for the unseen class during inference.

The training process spans 10 epochs, with 500 episodes per epoch for the UNSW-NB15 dataset and 2,000 episodes per epoch for the CIC-IoT23 dataset, reflecting their respective sizes (over 8,000 and 40,000 samples). Following training, the model is evaluated using 1,000 newly sampled episodes, structured similarly to the training episodes. To maintain a fair evaluation, training and testing episodes are drawn from disjoint subsets of the dataset. To further validate the robustness of our approach, we repeat the entire training and evaluation process over five independent iterations. For each iteration, the model’s weights are re-initialized, and the training procedure (including all epochs and episodes) is executed from scratch.

\begin{figure}[t!]
  \centering
  \includegraphics[width=\columnwidth]{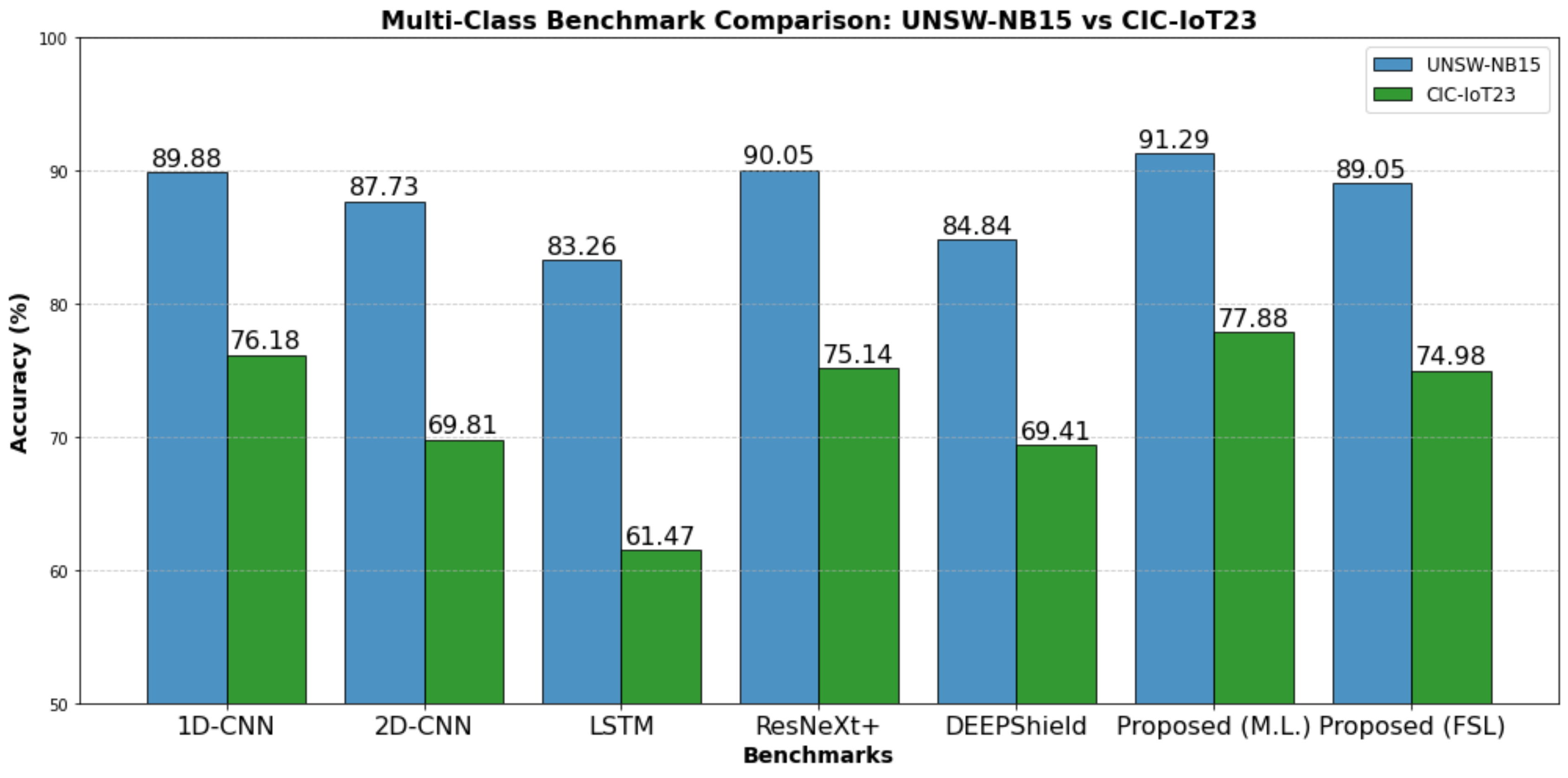}
  \caption{Comparison of accuracies of benchmarks vs. few-shot classification.}
  \label{fig:FewShotAcc}
\end{figure}

\subsubsection{Results}
The results of our few-shot learning experiments for classifying both known and newly introduced malware types are presented in Table \ref{tab:few-shot}. The \textit{Trained Classes} column indicates which classes were pre-trained with self-supervision, with the \textit{Unseen Class} column serving as the novel, untrained class for that specific experiment. In the 1-shot setting on the UNSW-NB15 dataset, the model's highest performance occurs when trained on Exploits and Generic malware types, with an overall query accuracy of 83.04\% and 82.82\% accuracy specifically for the unseen class. In the 5-shot setting, the best overall query accuracy of 89.48\% is achieved when the model is trained on Fuzzers and Generic malware types, with Exploits as the unseen class. Meanwhile, the highest accuracy for an unseen class is observed for the Generic class, achieving 89.95\%. For the CIC-IoT23 dataset, the model demonstrates a similar trend but with slightly lower accuracies, reflecting the increased diversity of the IoT traffic data. In the 1-shot setting, the best performance is achieved when training on Backdoor and Brute Force malware types, with Vulnerability as the unseen class, resulting in an overall query accuracy of 62.35\% and 61.73\% accuracy for the unseen class. Similarly, in the 5-shot setting, the model achieves its highest query accuracy of 75.01\% when trained on Backdoor and Brute Force malware types, with Vulnerability as the unseen class. 

\begin{figure}[t!]
    \centering
    \includegraphics[width=\columnwidth]{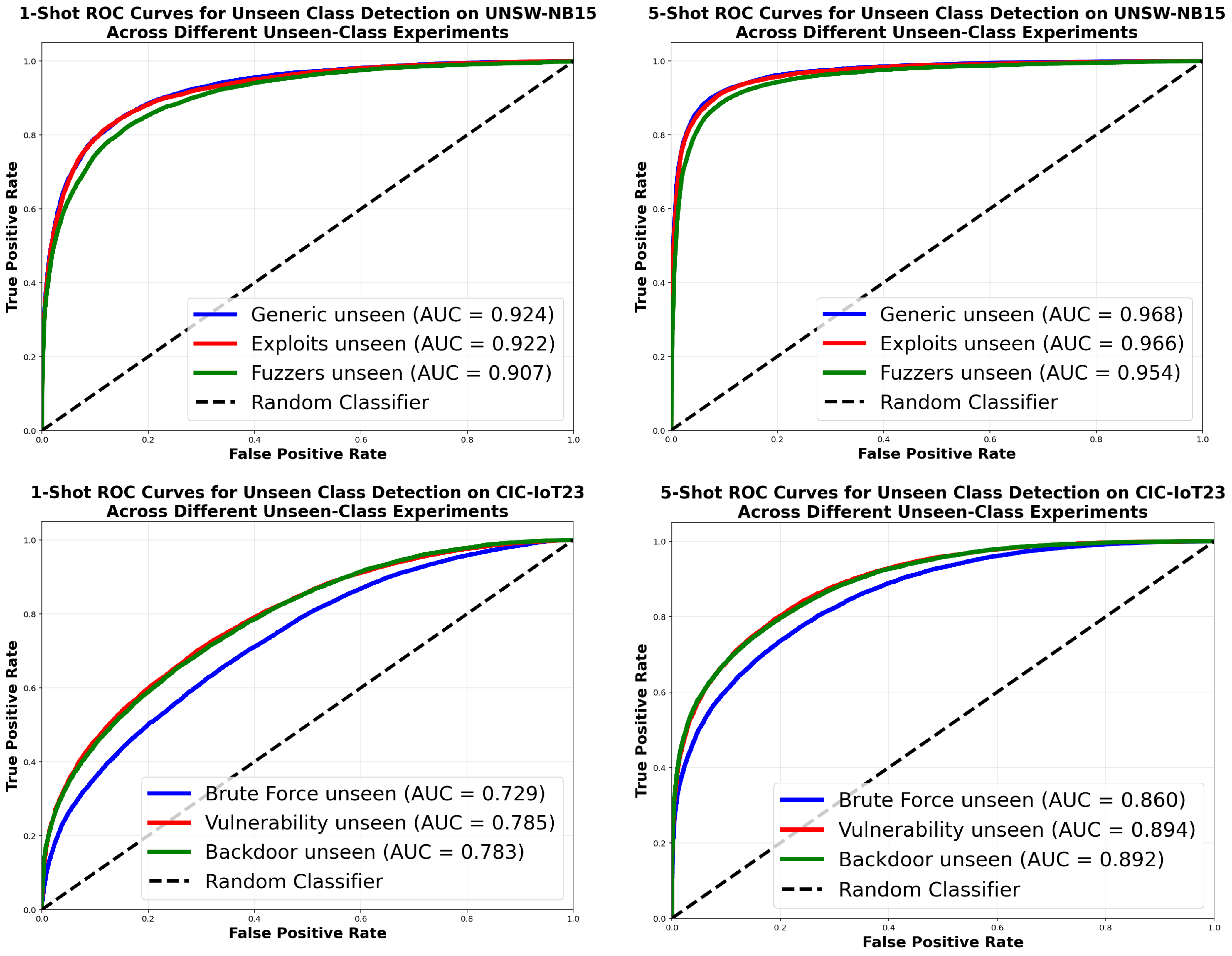}
    \caption{Few-Shot ROC for Unseen Class on UNSW-NB15 and CIC-IoT23, 1-/5-shot.} 
    \label{fig:ROC}
\end{figure}

Fig. \ref{fig:AccInf} offers additional context by comparing the relationship between accuracy and inference time for both datasets across different shot settings. The most significant accuracy improvement is observed between the 1-shot and 5-shot scenarios, with smaller incremental gains as the number of shots increases. The increase in accuracy between 1 and 5 shots shows how providing more support examples leads to the generation of more discriminative class prototypes. Specifically, on the UNSW-NB15 dataset, the model demonstrates a steady increase in accuracy with more shots before plateauing, showing optimal performance with minimal inference time. For example, the Generic class reaches approximately 90\% accuracy in the 15-shot setting, with an inference time of just over 2ms per episode. Similarly, in the CIC-IoT23 dataset, accuracy trends improve with more shots, with Vulnerability reaching over 76\% accuracy in the 15-shot setting. The inference times remain low for both datasets, demonstrating the computational efficiency of the proposed approach. Furthermore, it is important to note that an episode in our few-shot learning framework is not a single sample but consists of multiple samples: a support set containing $N_S$ labeled examples per class and a query set containing $N_Q$ unlabeled examples from all classes. For a 3-way classification task, each episode therefore includes \(3 \times (N_S + N_Q)\) samples.

Furthermore, we opt to compare the average few-shot results with the benchmarks to analyze how the accuracy compares, shown in Fig. \ref{fig:FewShotAcc}. Despite using significantly fewer training samples, the FSL method performs comparably to SOA classifiers, surpassing methods like 2D-CNN, LSTM, and DEEPShield classifiers. Additionally, in our FSL framework, we train on only two of three classes, unlike the previously mentioned multiclass classification problem, which required all three classes to undergo self-supervised training. This in-turn saves training costs and allows for the ability to detect previously unseen malware classes with high accuracy, an issue that is overlooked in typical classification problems. Overall, our approach effectively leverages knowledge from known malware types to classify novel malware types that were unseen during training, even with a limited number of labeled samples. This capability makes it a promising solution for real-world scenarios where labeled data is scarce and new malware types are frequent. 

Lastly, as shown in Fig. 6, we present the Receiver Operating Characteristic (ROC) curves corresponding to the unseen-class few-shot experiments. Each curve demonstrates model performance in both the 1-shot and 5-shot settings, reported separately for the UNSW-NB15 and CIC-IoT23 datasets. By plotting the true positive rate against the false positive rate, these curves highlight the model’s ability to distinguish unseen classes. For UNSW-NB15, all unseen classes achieve strong separability, with AUC values above 0.90 in both 1-shot and 5-shot settings. This indicates that even with minimal labeled support data, the learned embeddings provide highly discriminative features, and performance further improves with additional support examples. For CIC-IoT23, unseen detection is more challenging, as reflected in lower AUC values (0.73–0.78) in the 1-shot setting. However, the curves show improvement with 5-shots, with AUC values improving above 0.86 across all unseen categories. This demonstrates that the model generalizes better when provided with slightly more support samples, particularly in IoT traffic where class distributions and payload patterns are more diverse.

\begin{table}[t!]
\centering
\caption{Masking Percentage Ablation with M.L. Classifier (Binary)}
\label{tab:masking_perc}
\setlength{\tabcolsep}{6pt}
\renewcommand{\arraystretch}{1.2}
\scriptsize
\begin{tabular}{@{}lcccc@{}}
\toprule
\multirow{2}{*}{\textbf{Masking \%}} 
& \multicolumn{2}{c}{\textbf{UNSW-NB15}} 
& \multicolumn{2}{c}{\textbf{CIC-IoT23}} \\
\cmidrule(lr){2-3} 
\cmidrule(lr){4-5}
& \textbf{Accuracy} & \textbf{F1-Score} 
& \textbf{Accuracy} & \textbf{F1-Score} \\ 
\midrule
5\%   & \textbf{95.59 $\pm$ 0.22} & \textbf{95.59 $\pm$ 0.23} & {79.53 $\pm$ 0.33} & {79.48 $\pm$ 0.31}\\ 
10\%  & {95.47 $\pm$ 0.60} & {95.47 $\pm$ 0.60} & {80.69 $\pm$ 0.37} &{80.68 $\pm$ 0.37}\\ 
15\%  & {94.88 $\pm$ 0.15} & {94.88 $\pm$ 0.15} & {83.25 $\pm$ 0.25} & {82.34 $\pm$ 0.24}\\ 
20\%  & {94.28 $\pm$ 0.31} & {94.28 $\pm$ 0.31} & {83.17 $\pm$ 0.20} & { 83.15 $\pm$ 0.20 } \\ 
25\% & {94.57 $\pm$ 0.23} & {94.57 $\pm$ 0.23}  & \textbf{85.27 $\pm$ 0.21}  & \textbf{85.28 $\pm$ 0.21}  \\ 
30\%  & {93.97 $\pm$ 0.36} & {93.97 $\pm$ 0.36} & {84.93 $\pm$ 0.28} & {84.92 $\pm$ 0.28}\\ 
\bottomrule
\end{tabular}
\vspace{2mm}
\end{table}

\section{Ablation Study}

\subsection{Masking Percentage}

To investigate how varying proportion of masked bytes influences the quality of the learned embeddings, we conduct an ablation study examining multiple masking percentages ranging from 5\% to 30\%. Although 15\% masking is generally used for SSL tasks, our objective is to learn the optimal ratio that preserves sufficient information for robust training of the model, but also prevents the model from becoming too reliant on unmasked data. Table \ref{tab:masking_perc} displays the results for different masking percentages on distinguishing between malicious and benign packets. On the UNSW-NB15 dataset, we can observe the highest accuracies occur between 5\% and 15\%, with accuracy slightly dropping off when masking more than 15\% of packet bytes. On the more challenging CIC-IoT23 dataset, we observe that higher masking rates improve performance, with 25\% byte masking resulting in over a 2\% increase in classification over standard 15\% masking. 

The results suggest that the conventional 15\% of input masking commonly employed in SSL tasks may not be universally optimal for all network traffic domains. Instead, choosing an appropriate masking rate depends on dataset complexity and diversity. On more homogeneous network traffic, such as UNSW-NB15, a moderate masking strategy works well. However, on the CIC-IoT23 dataset, more varied patterns emerge due to complexity of the number of devices, over 100, connected on the same network. These additional complexities benefit from a higher masking ratio, which forces the model to rely more heavily on context rather than a smaller subset of repetitive byte patterns.

\subsection{Few-Shot Distance Metric}
To evaluate the impact on various distance metrics on few-shot classification, we experiment with three common measures: Euclidean distance, Cosine similarity, and Manhattan distance. Euclidean distance is used to measure the straight-line distance between the class prototype and the query point to be classified, cosine similarity is the cosine of the angle between two vectors and the dot product of the vectors normalized by the product of their lengths, and lastly Manhattan distance is computed as the sum of the absolute differences between coordinate values. Euclidean distance was exclusively implemented in our main experiments. By only changing the distance metric while keeping all other hyperparameters fixed, we can isolate the effect of the distance function on few-shot performance. 

\begin{figure}[t!]
  \centering
  \includegraphics[width=\columnwidth]{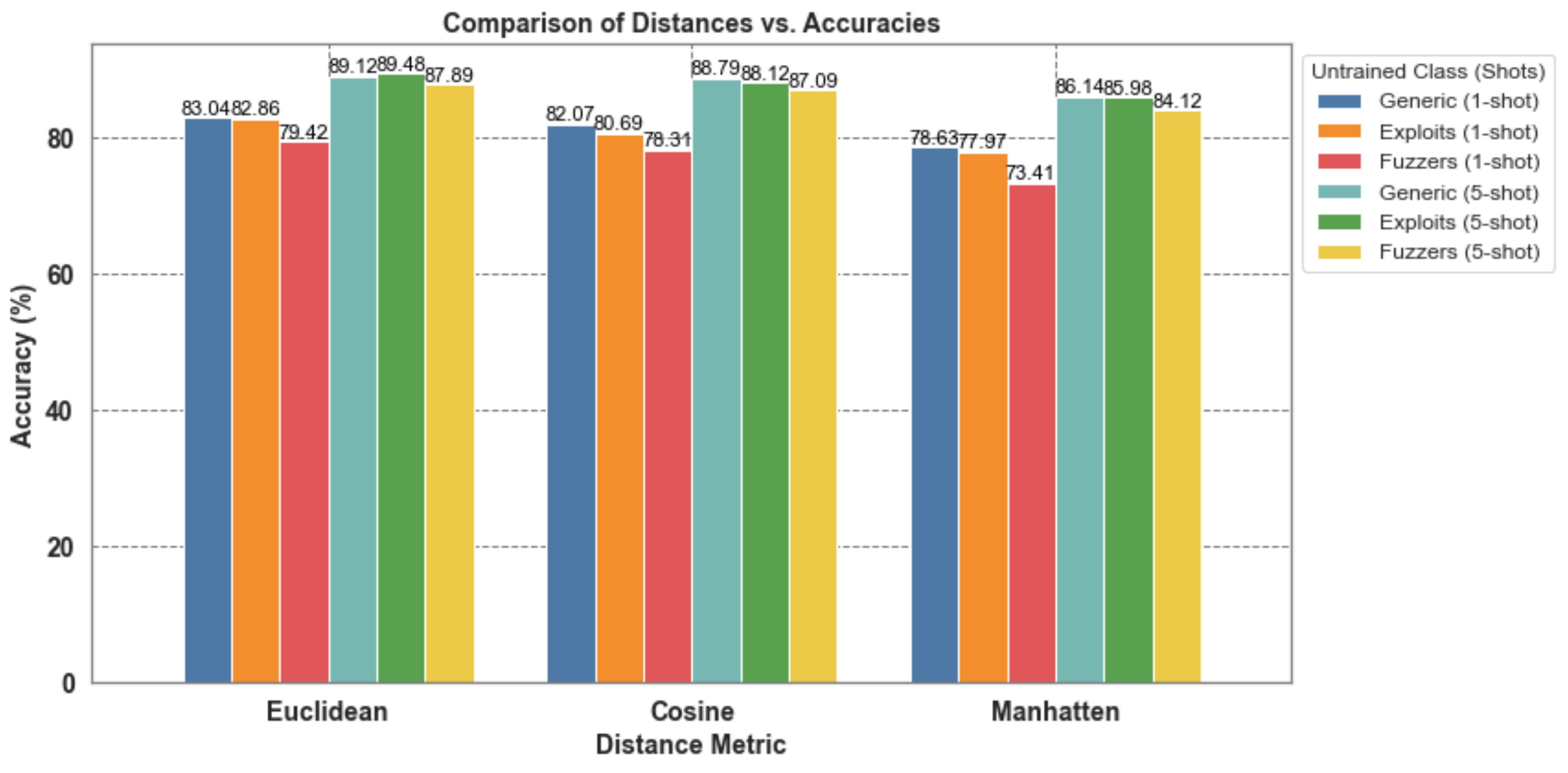} 
  \caption{Comparison of average few-shot classification accuracies across different distance metrics with the UNSW-NB15 dataset.}
  \label{fig:Distance_Fig} 
\end{figure}

The results of the ablation study are shown in  Fig. \ref{fig:Distance_Fig} and demonstrate notable variations in classification performance across all three distance metrics. Notably, Euclidean distance consistently outperforms Cosine and Manhattan distances on all three few-shot experiments with the UNSW-NB15 dataset, resulting in an average accuracy of 81.77\% $\pm$ 2.04\% and 88.83\% $\pm$ 0.83\% in the 1 and 5-shot settings, respectively. Cosine distance, although slightly less effective than Euclidean distance, still produces competitive results, achieving an average accuracy of 80.36\% $\pm$ 1.90\% in the 1-shot setting and 88.00\% $\pm$ 0.86\% in the 5-shot setting. Lastly, Manhattan distance demonstrates the weakest performance among the three distances, with an average accuracy of 76.67\% $\pm$ 2.84\% and 85.41\% $\pm$ 1.12\% in the 1 and 5-shot settings. 

\subsection{Sensitivity to Transformer Parameters}

To investigate the sensitivity to transformer depth and embedding dimension, we conduct an ablation study to evaluate the sensitivity of our model to the depth of the transformer backbone and the dimensionality of its embeddings. Our original architecture is based on a 12-layer transformer with a hidden size of 768, consistent with standard base configurations. To simulate resource-constrained settings for SSL pre-training, we designed variants with reduced numbers of transformer layers and hidden sizes. 

\begin{table}[t!]
\centering
\caption{Transformer depth and dimension ablation on UNSW-NB15 for binary, multiclass, and FSL classification.}
\label{tab:depth_dim_ablation_unsw}
\setlength{\tabcolsep}{6pt}
\renewcommand{\arraystretch}{1.2}
\scriptsize
\begin{tabular}{@{}lcccccc@{}}
\toprule
\multirow{2}{*}{\textbf{Depth / Dim}} 
& \multicolumn{2}{c}{\textbf{Binary}} 
& \multicolumn{2}{c}{\textbf{Multiclass}} 
& \multicolumn{2}{c}{\textbf{FSL (5-shot)}} \\
\cmidrule(lr){2-3} \cmidrule(lr){4-5} \cmidrule(lr){6-7}
& \textbf{Acc.} & \textbf{F1} 
& \textbf{Acc.} & \textbf{F1} 
& \textbf{Acc.} & \textbf{Unseen Acc.} \\ 
\midrule
6 / 384             & 76.03 & 75.06 & 85.46 & 85.45 & 68.26 & 67.52 \\ 
6 / 768             & 92.57 & 92.57 & 89.31 & 89.31 & 82.88 & 82.60 \\ 
12 / 384             & 82.81 & 82.60 & 84.73 & 84.68 & 70.19 & 69.42 \\ 
\midrule
12 / 768 (baseline) & 94.88 & 94.88 & 91.29 & 92.29 & 88.83 & 88.93 \\ 
\bottomrule
\end{tabular}
\vspace{2mm}
\end{table}

The rationale behind this design consists of two main reasons. First, the number of transformer layers in the model influences the representational capacity of the model, while also corresponding to the number of attention heads the model can apply. This enables deeper networks to capture longer-range dependencies and better refine attention patterns, although at the cost of increased computational overhead during pre-training. Second, the size of the embedding dimension determines how much information is retained in the latent space. Smaller dimensions may improve computational efficiency, but lose fine-grained and discriminative characteristics of the traffic bytes. By evaluating different configurations, we can better understand the relationships between model depth, representational power, and classification performance across binary, multiclass, and FSL scenarios.

Table 8 summarizes the results of this ablation on the UNSW-NB15 dataset, compared to baseline classification performance of each experiment. We report accuracy and F1 metrics across binary and multiclass experiments, and overall and unseen class accuracies for the FSL experiment based on the 5-shot baseline results, averaged over all three FSL experiments. Reducing transformer layer depth from 12 to 6 layers while retaining the 768-dimensional embedding yields the smallest degradation across all tasks. In contrast, narrowing the embedding to 384 dimensions across either depth substantially decreases performance, with the largest drops occurring in the binary and FSL classification tasks. The baseline models remains best overall, but among the reduced-capacity models, we can conclude that the 6/768 variant offers a favorable trade-off.

This occurs due to the smaller embedding dimension constraining the model's capacity to encode discriminative features from the traffic data, especially under few-shot scenarios, where the model must adapt to the novel class that was not seen during pre-training. Consequently, this leads to the largest performance drop in the few-shot setting for average and unseen accuracies. Regarding the depth of the layers, we can conclude that most discriminative features are learned in the earlier layers of the model. However, the baseline performance of 12/768 still yields favorable results compared to the constrained variants.

\section{Encrypted Traffic}
\label{sec:threat}

In this study, our analysis focuses on the raw packet bytes contained within TCP and UDP network packets from various datasets. However, it is crucial to note the limitations when dealing with encrypted traffic.  Encryption---the process of converting plaintext into ciphertext to secure data transmission and make it unreadable to unauthorized entities without the appropriate decryption keys---plays a pivotal role in data security. Specifically, when plaintext is encrypted using a key, it transforms into ciphertext that does not reveal any information about the original content without the correct decryption key~\cite{menezes2018handbook}.

Let us denote \(m\) as the variable representing the raw payload message, and \(E_k(\cdot)\) as the encryption function that encrypts \(m\) with the key \(k\). In this context, observing the ciphertext \(c\) yields no information about the original message \(m\), expressed as:
\begin{equation}
    p(m = m_1 | E_k(m) = c) = p
    \label{eq:encryptionequation}
\end{equation}
This equation underscores that decrypting the plaintext \(m_1\) with the encryption algorithm \(E_k(\cdot)\) and key \(k\) results in a unique ciphertext, effectively concealing any details of the plaintext.

\begin{figure}[t!]
  \centering
  \includegraphics[width=1.0\linewidth]{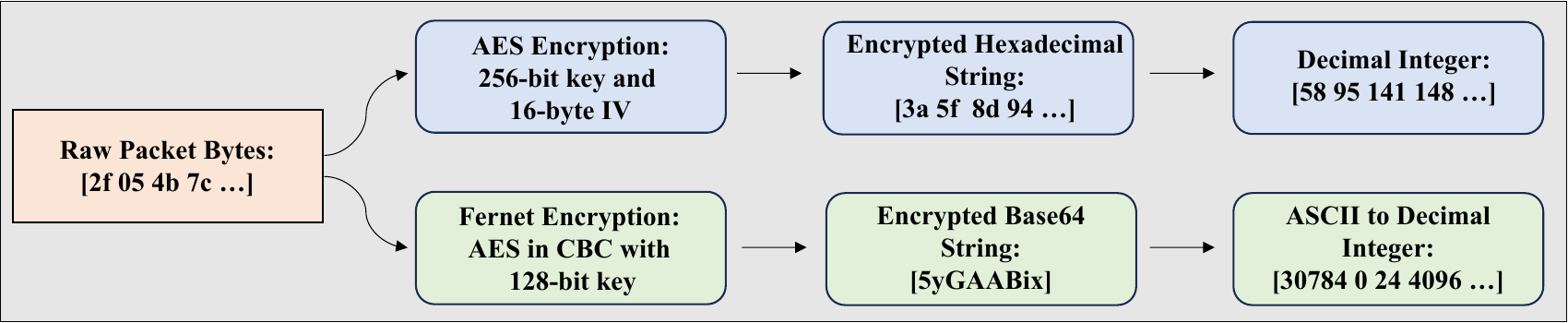}
  \caption{Comparison of processing input packets for AES and Fernet encryption processes.}
  \label{fig:Encryption} 
\end{figure}

Despite these encryption properties, some studies have successfully classified encrypted network traffic from various applications \cite{ socialmedia, deeppacket}. This classification success is largely due to the unique encryption signatures that different applications' random generators leave, which, while not disclosing plaintext content, offer a form of application-specific pattern. The challenge, however, lies in malware detection, where access to plaintext information is crucial. If malware signatures can be detected in encrypted traffic, it would suggest that the encryption method fails to adequately mask the plaintext, thereby violating the principle stated in Equation \ref{eq:encryptionequation}. Our research involved applying two encryption methods to the packet data: the Advanced Encryption Standard (AES) and Fernet symmetric encryption \cite{pycryptodome_aes, cryptographyio_fernet}. Unlike our earlier work \cite{stein2024towards}, which introduced text-encoding artifacts through character-level tokenization, this byte-level formulation avoids dataset-specific cues and provides a more rigorous basis for evaluating ciphertext learnability. Fig. \ref{fig:Encryption} represents the different pre-processing methods between both encryption methods. 

\subsection{AES Encryption}
AES encryption uses a 256-bit key and a 16-byte initialization vector (IV), ensuring a unique encryption outcome for each key and IV pair. The uniqueness of AES lies in its ability to obscure any recognizable patterns within the payload data by yielding an unrepeatable outcome for every unique combination of key and IV. This guarantees that even if the same packet data is encrypted multiple times, the outcomes will differ each time when a different IV is used. After encryption, the AES-encrypted packets are converted into a hexadecimal format. Each character in this hexadecimal representation is then mapped to its corresponding decimal integer, forming a sequence that is padded to uniform sequence lengths across the dataset. The M.L. model's average test accuracy with AES-encrypted data was 50.40\% with an F1-Score of 50.40\%. Similar results were observed when testing other previously mentioned SOA methods, none of which achieved an accuracy greater than 50\% in classifying malicious or benign packets. These metrics indicate that AES encryption substantially diminishes the model's ability to discern between malicious and benign payloads.

\subsection{Fernet Encryption}
Fernet employs AES in Cipher Block Chaining (CBC) mode with a 128-bit key. Fernet's encrypted outputs are base64-encoded, resulting in a more diverse character set. Similar to AES, the encrypted payloads under Fernet encryption are also converted into sequences of decimal integers. However, due to the base64 encoding, each character is first converted to its ASCII value before the sequence is formed. This process allows for a larger vocabulary size since base64 encoding includes a broader range of characters. The application of Fernet encryption resulted in test accuracy of 51.64\% and an F1-Score of 51.57\%. Once again, when tested with other SOA classifiers, the model fails to distinguish malicious from benign content with better than random guessing performance. This suggests Fernet effectively conceals patterns within the data by pairing strong AES encryption with Base64 encoding, therefore leaving deep-learning methods with no contextual cues for reliable detection.

\section{Discussion and Limitations}
\label{sec:discussion}
\vspace{2mm}
\subsection{Discussion}
Our research stands out in several key ways. We began by removing duplicated packet byte values, enhancing the model's exposure to unique packets and thereby improving data quality for more effective generalization to the testing dataset. Retaining these duplicates can reduce the uniqueness of signatures, making the process for classification algorithms easier to identify which packets belong to a certain class. Another pivotal aspect of our approach is the strict focus on packet bytes as inputs. Previous classification methods typically rely on statistical-based feature methodologies, such as size of the packet, port numbers, flow duration, etc., derived from traffic data. While this approach is sufficient for traditional machine learning algorithms, sequence-based inputs, such as byte sequences, benefit significantly more from modeling long-term dependencies. Our analysis was also not limited to the initial 784 payload bytes as described in \cite{b9}, but extended to the MTU of 1500 bytes. Given that the typical MTU for Ethernet packet-based networks is 1500 bytes, with around 40 bytes allocated for header information, this approach ensures our model considers the full length of available payload data. This is important as malicious content might be embedded further within a packet's payload. By removing the constraint of shortening the byte data, we address the broader complexity within in each payload.

In this study, it's important to acknowledge the distinct types of data used in each dataset, which reflect different network scenarios. The CIC-IOT23 dataset is comprised of network traffic from 105 devices within an IoT network topology, designed to capture traffic patterns typical of IoT environments. However, the UNSW-NB15 dataset was constructed to provide a mix of benign activities and synthetic malicious behaviors, aiming to support a broad spectrum of network intrusion detection research. This dataset encompasses a variety of attack types and network behaviors, making it structurally and contextually different from the CIC-IOT23. The variation in network-connected devices across the datasets not only influences the nature of the network traffic but also highlights the unique operational scenarios each dataset is intended to simulate. Furthermore, the CIC-IOT23 dataset likely presents unique challenges and patterns that are not as prevalent in the UNSW-NB15 dataset. 

An additional observation is that the simple ML classifier consistently outperforms more complex learning architectures such as CNNs and LSTMs when leveraging the SSL embeddings, despite its relatively lightweight structure. This can be explained by the strength of the SSL-generated embeddings, which already encode local and global dependencies within packet data. Since these embeddings are highly discriminative, the ML classifier can leverage them directly without requiring additional temporal feature extraction. In contrast, the deeper models may introduce unnecessary complexity or risk overfitting on the SSL embeddings. This finding demonstrates that once robust embeddings are learned, a simple downstream classifier can lead to improved accuracy and faster inference, which is especially beneficial for real-time deployment.

The practical implementation entails deploying the malware detection model onto edge devices within the network layer. The edge devices are responsible for connecting different segments of the network, enabling traffic between networks through a physical component. The malware detection model would be implemented directly on these devices, allowing for the immediate analysis of network traffic as it passes through. This placement means that the model can efficiently filter and examine data packets in real-time, identifying potential threats at the network's edge. Also, this model could be deployed across numerous edge devices to provide comprehensive coverage of a network. 

\subsection{Limitations and Threats to Validity}
Our study demonstrates promising results, but several limitations and threats to validity should be acknowledged. Pre-training with transformers is computationally expensive and depends heavily on dataset scale \cite{beal2022billion, gui2024survey}. Larger and more diverse datasets generally yield richer embeddings but require substantial compute and training time, which may limit accessibility for organizations without high-performance infrastructure. However, the growing availability of high-performance GPUs is making large-scale pre-training on datasets of benign and malicious behaviors increasingly feasible. In addition, because our SSL embeddings are fixed after pre-training, downstream tasks are constrained by the representations learned from the pre-training dataset. This can limit the model’s ability to continuously generalize to novel malware families over time. While our proposed FSL method partially mitigates this, future work could investigate incremental or online SSL training to refine embeddings in real time. With parameter-efficient fine-tuning methods such as LoRA \cite{hu2022lora}), this becomes increasingly practical, since only a small set of additive parameters must be updated rather than fully fine-tuning the entire model.

Furthermore, our approach assumes access to plaintext transport-layer payloads for DPI. In environments with widespread encryption, payload observability is reduced, and our model performs at chance when packet bytes are fully encrypted using Fernet or AES. We note earlier in the manuscript that DPI often presumes access to unencrypted payloads \cite{deri2021using, dijk2021detection}, a condition that can still hold in certain enterprise networks, industrial control systems (ICS), and CAN applications. Future work will combine payload-agnostic statistics with our SSL embeddings and evaluate on further encrypted captures.

We trained and evaluated on two public corpora (UNSW-NB15 and CIC-IoT23) with deliberate class selection and de-duplication to balance classes; nonetheless, focusing on a limited set of attack families per corpus and sampling subsets can introduce bias in byte distributions and background traffic characteristics. While performance and few-shot transfer indicate promise within the chosen attacks across these two datasets, external validity remains limited to these settings. To strengthen generalization claims, future work will conduct cross-generalization capabilities of the proposed model, incorporate additional captures spanning network traffic collection pipelines, and evaluate domain-adaptation and deployability strategies for out-of-distribution real-time malware detection.

\section{Conclusion and Future Research Directions}
\label{sec:conclusion}

This paper explored the application of a self-supervised transformer-based model with a downstream task and few-shot learning for malware detection and classification. The model was evaluated on the UNSW-NB15 and CIC-IOT23 datasets, focusing on the packets of UDP and TCP packets serving as inputs. Our method showed robust results on the classification of benign versus malicious packets when compared to SOA methods on both datasets. Furthermore, our results show that a simple ML classifier, when paired with SSL-generated embeddings, can outperform more complex CNN and LSTM models. This emphasizes that SSL pre-training results in discriminative embeddings, allowing lightweight models to achieve improved accuracy and efficiency compared to deeper architectures, which is advantageous for real-time intrusion detection. The few-shot results were also promising, demonstrating the model's ability to learn with limited amounts of labeled data. Using the packet bytes as input vectors, we were able to classify when different packets are either benign or malicious, as well as which types of attacks were present. While the payload bytes of packets are inherently different from the natural language structure, this study shows that the transformer-based model developed for natural language can be leveraged to capture and learn the intricate sequential patterns of the packet bytes.

Despite these encouraging results, several challenges and opportunities remain. Although our evaluation focused on traditional enterprise network and  IoT packets, it is crucial to extend out work to additional protocols across several domains. In particular, protocols associated with ICS and in-vehicle CAN present unique characteristics which may require additional pre-training. Furthermore, another promising direction is the design and implementation of new models based off our proposed transformer-based architecture that incorporates emerging paradigms such as continual learning, which would enable rapid adaptation to evolving traffic patterns while remaining suitable for deployment in resource-constrained environments. Lastly, building attribution models that leverage distinctive network packet signatures represents an important challenge, as such models would enhance interpretability by linking malicious behaviors to identifiable network traits. Overall, this work proposed a self-supervised transformer-based pre-training framework for malware detection and classification, demonstrating its versatility through multiple downstream tasks, including FSL.

\bibliographystyle{IEEEtran} 
\bibliography{main}

\begin{thebibliography}{10}
\providecommand{\url}[1]{#1}
\csname url@samestyle\endcsname
\providecommand{\newblock}{\relax}
\providecommand{\bibinfo}[2]{#2}
\providecommand{\BIBentrySTDinterwordspacing}{\spaceskip=0pt\relax}
\providecommand{\BIBentryALTinterwordstretchfactor}{4}
\providecommand{\BIBentryALTinterwordspacing}{\spaceskip=\fontdimen2\font plus
\BIBentryALTinterwordstretchfactor\fontdimen3\font minus
  \fontdimen4\font\relax}
\providecommand{\BIBforeignlanguage}[2]{{%
\expandafter\ifx\csname l@#1\endcsname\relax
\typeout{** WARNING: IEEEtran.bst: No hyphenation pattern has been}%
\typeout{** loaded for the language `#1'. Using the pattern for}%
\typeout{** the default language instead.}%
\else
\language=\csname l@#1\endcsname
\fi
#2}}
\providecommand{\BIBdecl}{\relax}
\BIBdecl

\bibitem{firewall}
H.~J. Kiratsata, D.~P. Raval, P.~K. Viras, P.~Lalwani, H.~Patel, and
  S.~Panchal, ``Behaviour analysis of open-source firewalls under security
  crisis,'' in \emph{2022 International Conference on Wireless Communications
  Signal Processing and Networking (WiSPNET)}.\hskip 1em plus 0.5em minus
  0.4em\relax IEEE, 2022, pp. 105--109.

\bibitem{antivirus}
C.~Rohith and G.~Kaur, ``A comprehensive study on malware detection and
  prevention techniques used by anti-virus,'' in \emph{2021 2nd international
  conference on intelligent engineering and management (iciem)}.\hskip 1em plus
  0.5em minus 0.4em\relax IEEE, 2021, pp. 429--434.

\bibitem{DPI1}
O.~N. Nyasore, P.~Zavarsky, B.~Swar, R.~Naiyeju, and S.~Dabra, ``Deep packet
  inspection in industrial automation control system to mitigate attacks
  exploiting modbus/tcp vulnerabilities,'' in \emph{2020 IEEE 6th Intl
  Conference on Big Data Security on Cloud (BigDataSecurity), IEEE Intl
  Conference on High Performance and Smart Computing,(HPSC) and IEEE Intl
  Conference on Intelligent Data and Security (IDS)}.\hskip 1em plus 0.5em
  minus 0.4em\relax IEEE, 2020, pp. 241--245.

\bibitem{DPI2}
G.~Aceto, D.~Ciuonzo, A.~Montieri, and A.~Pescap{\'e}, ``Mobile encrypted
  traffic classification using deep learning: Experimental evaluation, lessons
  learned, and challenges,'' \emph{IEEE transactions on network and service
  management}, vol.~16, no.~2, pp. 445--458, 2019.

\bibitem{b1}
J.~Devlin, ``Bert: Pre-training of deep bidirectional transformers for language
  understanding,'' \emph{arXiv preprint arXiv:1810.04805}, 2018.

\bibitem{he2016deep}
K.~He, X.~Zhang, S.~Ren, and J.~Sun, ``Deep residual learning for image
  recognition,'' in \emph{Proceedings of the IEEE conference on computer vision
  and pattern recognition}, 2016, pp. 770--778.

\bibitem{Attention}
A.~Vaswani, ``Attention is all you need,'' \emph{Advances in Neural Information
  Processing Systems}, 2017.

\bibitem{malwarestats}
\BIBentryALTinterwordspacing
``Malware statistics,” dataprot. [online]. available,'' last accessed 20Jan
  2024. [Online]. Available:
  \url{https://dataprot.net/statistics/malware-statistics/}
\BIBentrySTDinterwordspacing

\bibitem{b3}
G.~Aceto, D.~Ciuonzo, A.~Montieri, and A.~Pescap{\'e}, ``Mobile encrypted
  traffic classification using deep learning: Experimental evaluation, lessons
  learned, and challenges,'' \emph{IEEE transactions on network and service
  management}, vol.~16, no.~2, pp. 445--458, 2019.

\bibitem{Patheja}
A.~Javaid, Q.~Niyaz, W.~Sun, and M.~Alam, ``A deep learning approach for
  network intrusion detection system,'' in \emph{Proceedings of the 9th EAI
  International Conference on Bio-inspired Information and Communications
  Technologies (formerly BIONETICS)}, 2016, pp. 21--26.

\bibitem{DDOS}
R.~Doshi, N.~Apthorpe, and N.~Feamster, ``Machine learning ddos detection for
  consumer internet of things devices,'' in \emph{2018 IEEE Security and
  Privacy Workshops (SPW)}.\hskip 1em plus 0.5em minus 0.4em\relax IEEE, 2018,
  pp. 29--35.

\bibitem{IOT}
J.~G. Almaraz-Rivera, J.~A. Perez-Diaz, and J.~A. Cantoral-Ceballos,
  ``Transport and application layer ddos attacks detection to iot devices by
  using machine learning and deep learning models,'' \emph{Sensors}, vol.~22,
  no.~9, p. 3367, 2022.

\bibitem{CANRnn}
K.~Stein, A.~Mahyari, and E.~El-Sheikh, ``Vehicle controller area network
  inspection using recurrent neural networks,'' in \emph{International
  Conference on Advances in Computing Research}.\hskip 1em plus 0.5em minus
  0.4em\relax Springer, 2023, pp. 494--499.

\bibitem{saiyed2024deep}
M.~F. Saiyed and I.~Al-Anbagi, ``Deep ensemble learning with pruning for ddos
  attack detection in iot networks,'' \emph{IEEE Transactions on Machine
  Learning in Communications and Networking}, 2024.

\bibitem{transaction}
Y.~He, X.~Kang, Q.~Yan, and E.~Li, ``Resnext+: Attention mechanisms based on
  resnext for malware detection and classification,'' \emph{IEEE Transactions
  on Information Forensics and Security}, 2023.

\bibitem{Wf-transformer}
Q.~Zhou, L.~Wang, H.~Zhu, T.~Lu, and V.~S. Sheng, ``Wf-transformer: Learning
  temporal features for accurate anonymous traffic identification by using
  transformer networks,'' \emph{IEEE Transactions on Information Forensics and
  Security}, 2023.

\bibitem{ravi2023attention}
V.~Ravi and M.~Alazab, ``Attention-based convolutional neural network deep
  learning approach for robust malware classification,'' \emph{Computational
  Intelligence}, vol.~39, no.~1, pp. 145--168, 2023.

\bibitem{yu2020intrusion}
Y.~Yu and N.~Bian, ``An intrusion detection method using few-shot learning,''
  \emph{IEEE Access}, vol.~8, pp. 49\,730--49\,740, 2020.

\bibitem{lu2023few}
C.~Lu, X.~Wang, A.~Yang, Y.~Liu, and Z.~Dong, ``A few-shot-based model-agnostic
  meta-learning for intrusion detection in security of internet of things,''
  \emph{IEEE Internet of Things Journal}, vol.~10, no.~24, pp.
  21\,309--21\,321, 2023.

\bibitem{zhou2020siamese}
X.~Zhou, W.~Liang, S.~Shimizu, J.~Ma, and Q.~Jin, ``Siamese neural network
  based few-shot learning for anomaly detection in industrial cyber-physical
  systems,'' \emph{IEEE Transactions on Industrial Informatics}, vol.~17,
  no.~8, pp. 5790--5798, 2020.

\bibitem{stein2024towards}
K.~Stein, A.~A. Mahyari, G.~Francia, and E.~El-Sheikh, ``Towards novel
  malicious packet recognition: A few-shot learning approach,'' in \emph{MILCOM
  2024-2024 IEEE Military Communications Conference (MILCOM)}.\hskip 1em plus
  0.5em minus 0.4em\relax IEEE, 2024, pp. 847--852.

\bibitem{b9}
M.~J. De~Lucia, P.~E. Maxwell, N.~D. Bastian, A.~Swami, B.~Jalaian, and
  N.~Leslie, ``Machine learning raw network traffic detection,'' in
  \emph{Artificial Intelligence and Machine Learning for Multi-Domain
  Operations Applications III}, vol. 11746.\hskip 1em plus 0.5em minus
  0.4em\relax SPIE, 2021, pp. 185--194.

\bibitem{b10}
L.~Xu, X.~Zhou, Y.~Ren, and Y.~Qin, ``A traffic classification method based on
  packet transport layer payload by ensemble learning,'' in \emph{2019 IEEE
  Symposium on Computers and Communications (ISCC)}.\hskip 1em plus 0.5em minus
  0.4em\relax IEEE, 2019, pp. 1--6.

\bibitem{SSL}
J.~Gui, T.~Chen, J.~Zhang, Q.~Cao, Z.~Sun, H.~Luo, and D.~Tao, ``A survey on
  self-supervised learning: Algorithms, applications, and future trends,''
  \emph{arXiv preprint arXiv:2301.05712}, 2023.

\bibitem{SSL2}
X.~Liu, F.~Zhang, Z.~Hou, L.~Mian, Z.~Wang, J.~Zhang, and J.~Tang,
  ``Self-supervised learning: Generative or contrastive,'' \emph{IEEE
  transactions on knowledge and data engineering}, vol.~35, no.~1, pp.
  857--876, 2021.

\bibitem{Network}
A.~Gupta, A.~Raj, M.~Arora \emph{et~al.}, ``Ip traffic classification of 4g
  network using machine learning techniques,'' in \emph{2021 5th International
  conference on computing methodologies and communication (ICCMC)}.\hskip 1em
  plus 0.5em minus 0.4em\relax IEEE, 2021, pp. 127--132.

\bibitem{zhang2020ransomware}
B.~Zhang, W.~Xiao, X.~Xiao, A.~K. Sangaiah, W.~Zhang, and J.~Zhang,
  ``Ransomware classification using patch-based cnn and self-attention network
  on embedded n-grams of opcodes,'' \emph{Future Generation Computer Systems},
  vol. 110, pp. 708--720, 2020.

\bibitem{TCPvsUDP}
\BIBentryALTinterwordspacing
A.~Andriekutė. (2023) Tcp vs udp: What’s the main difference? Accessed:
  [Insert Access Date]. [Online]. Available:
  \url{https://nordvpn.com/blog/tcp-or-udp-which-is-better/}
\BIBentrySTDinterwordspacing

\bibitem{kumar2012survey}
S.~Kumar and S.~Rai, ``Survey on transport layer protocols: Tcp \& udp,''
  \emph{International Journal of Computer Applications}, vol.~46, no.~7, pp.
  20--25, 2012.

\bibitem{farrukh2022payload}
Y.~A. Farrukh, I.~Khan, S.~Wali, D.~Bierbrauer, J.~A. Pavlik, and N.~D.
  Bastian, ``Payload-byte: A tool for extracting and labeling packet capture
  files of modern network intrusion detection datasets,'' in \emph{2022
  IEEE/ACM International Conference on Big Data Computing, Applications and
  Technologies (BDCAT)}.\hskip 1em plus 0.5em minus 0.4em\relax IEEE, 2022, pp.
  58--67.

\bibitem{deri2021using}
L.~Deri and F.~Fusco, ``Using deep packet inspection in cybertraffic
  analysis,'' in \emph{2021 IEEE International Conference on Cyber Security and
  Resilience (CSR)}.\hskip 1em plus 0.5em minus 0.4em\relax IEEE, 2021, pp.
  89--94.

\bibitem{dijk2021detection}
A.~Dijk, ``Detection of advanced persistent threats using artificial
  intelligence for deep packet inspection,'' in \emph{2021 IEEE International
  Conference on Big Data (Big Data)}.\hskip 1em plus 0.5em minus 0.4em\relax
  IEEE, 2021, pp. 2092--2097.

\bibitem{OneShot}
O.~Vinyals, C.~Blundell, T.~Lillicrap, D.~Wierstra \emph{et~al.}, ``Matching
  networks for one shot learning,'' \emph{Advances in neural information
  processing systems}, vol.~29, 2016.

\bibitem{PrototypicalNetworks}
J.~Snell, K.~Swersky, and R.~Zemel, ``Prototypical networks for few-shot
  learning,'' \emph{Advances in neural information processing systems},
  vol.~30, 2017.

\bibitem{LaplacianNetworks}
I.~Ziko, J.~Dolz, E.~Granger, and I.~B. Ayed, ``Laplacian regularized few-shot
  learning,'' in \emph{International conference on machine learning}.\hskip 1em
  plus 0.5em minus 0.4em\relax PMLR, 2020, pp. 11\,660--11\,670.

\bibitem{stein2025transductive}
K.~Stein, A.~A. Mahyari, G.~Francia, and E.~El-Sheikh, ``Transductive one-shot
  learning meet subspace decomposition,'' in \emph{2025 IEEE International
  Conference on Image Processing (ICIP)}.\hskip 1em plus 0.5em minus
  0.4em\relax IEEE, 2025, pp. 576--581.

\bibitem{UNSW}
N.~Moustafa and J.~Slay, ``Unsw-nb15: a comprehensive data set for network
  intrusion detection systems (unsw-nb15 network data set),'' in \emph{2015
  military communications and information systems conference (MilCIS)}.\hskip
  1em plus 0.5em minus 0.4em\relax IEEE, 2015, pp. 1--6.

\bibitem{CICIOT}
E.~C.~P. Neto, S.~Dadkhah, R.~Ferreira, A.~Zohourian, R.~Lu, and A.~A.
  Ghorbani, ``Ciciot2023: A real-time dataset and benchmark for large-scale
  attacks in iot environment,'' \emph{Sensors}, vol.~23, no.~13, p. 5941, 2023.

\bibitem{KDD9}
M.~Tavallaee, E.~Bagheri, W.~Lu, and A.~A. Ghorbani, ``A detailed analysis of
  the kdd cup 99 data set,'' in \emph{2009 IEEE symposium on computational
  intelligence for security and defense applications}.\hskip 1em plus 0.5em
  minus 0.4em\relax IEEE, 2009, pp. 1--6.

\bibitem{kingma2017adammethodstochasticoptimization}
\BIBentryALTinterwordspacing
D.~P. Kingma and J.~Ba, ``Adam: A method for stochastic optimization,'' 2017.
  [Online]. Available: \url{https://arxiv.org/abs/1412.6980}
\BIBentrySTDinterwordspacing

\bibitem{huggingface_optimizer}
\BIBentryALTinterwordspacing
H.~Face, ``Main classes - optimizer and schedules,'' 2023, transformers
  Documentation, 2023. [Online]. Available. [Online]. Available:
  \url{https://huggingface.co/docs/transformers/main_classes/optimizer_schedules}
\BIBentrySTDinterwordspacing

\bibitem{kingma2014adam}
D.~P. Kingma, ``Adam: A method for stochastic optimization,'' \emph{arXiv
  preprint arXiv:1412.6980}, 2014.

\bibitem{episodic}
S.~Laenen and L.~Bertinetto, ``On episodes, prototypical networks, and few-shot
  learning,'' \emph{Advances in Neural Information Processing Systems},
  vol.~34, pp. 24\,581--24\,592, 2021.

\bibitem{ali2022effective}
S.~Ali, O.~Abusabha, F.~Ali, M.~Imran, and T.~Abuhmed, ``Effective multitask
  deep learning for iot malware detection and identification using behavioral
  traffic analysis,'' \emph{IEEE Transactions on Network and Service
  Management}, vol.~20, no.~2, pp. 1199--1209, 2022.

\bibitem{menezes2018handbook}
A.~J. Menezes, P.~C. Van~Oorschot, and S.~A. Vanstone, \emph{Handbook of
  applied cryptography}.\hskip 1em plus 0.5em minus 0.4em\relax CRC press,
  2018.

\bibitem{socialmedia}
F.~Al-Obaidy, S.~Momtahen, M.~F. Hossain, and F.~Mohammadi, ``Encrypted traffic
  classification based ml for identifying different social media
  applications,'' in \emph{2019 IEEE Canadian Conference of Electrical and
  Computer Engineering (CCECE)}.\hskip 1em plus 0.5em minus 0.4em\relax IEEE,
  2019, pp. 1--5.

\bibitem{deeppacket}
M.~Lotfollahi, M.~Jafari~Siavoshani, R.~Shirali Hossein~Zade, and M.~Saberian,
  ``Deep packet: A novel approach for encrypted traffic classification using
  deep learning,'' \emph{Soft Computing}, vol.~24, no.~3, pp. 1999--2012, 2020.

\bibitem{pycryptodome_aes}
\BIBentryALTinterwordspacing
PyCryptodome, ``Aes - advanced encryption standard,'' 2021, 2021. [Online].
  Available. [Online]. Available:
  \url{https://pycryptodome.readthedocs.io/en/latest/src/cipher/aes.html}
\BIBentrySTDinterwordspacing

\bibitem{cryptographyio_fernet}
\BIBentryALTinterwordspacing
Cryptography.io, ``Fernet (symmetric encryption),'' 2021, 2021. [Online].
  Available. [Online]. Available:
  \url{https://cryptography.io/en/latest/fernet/}
\BIBentrySTDinterwordspacing

\bibitem{beal2022billion}
J.~Beal, H.-Y. Wu, D.~H. Park, A.~Zhai, and D.~Kislyuk, ``Billion-scale
  pretraining with vision transformers for multi-task visual representations,''
  in \emph{Proceedings of the IEEE/CVF winter conference on applications of
  computer vision}, 2022, pp. 564--573.

\bibitem{gui2024survey}
J.~Gui, T.~Chen, J.~Zhang, Q.~Cao, Z.~Sun, H.~Luo, and D.~Tao, ``A survey on
  self-supervised learning: Algorithms, applications, and future trends,''
  \emph{IEEE Transactions on Pattern Analysis and Machine Intelligence},
  vol.~46, no.~12, pp. 9052--9071, 2024.

\bibitem{hu2022lora}
E.~J. Hu, Y.~Shen, P.~Wallis, Z.~Allen-Zhu, Y.~Li, S.~Wang, L.~Wang, W.~Chen
  \emph{et~al.}, ``Lora: Low-rank adaptation of large language models.''
  \emph{ICLR}, vol.~1, no.~2, p.~3, 2022.

\end{thebibliography}

\begin{IEEEbiography}[{\includegraphics[width=1in,height=1.25in,clip,keepaspectratio]{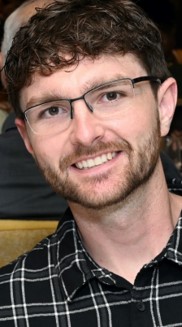}}]{Kyle Stein}
is a Ph.D. candidate in Intelligent Systems and Robotics at the University of West Florida, Pensacola, FL, supported by a CyberCorps Scholarship for Service (SFS). He received the M.S. degree in Mathematical Sciences from the University of West Florida, Pensacola, FL, in 2021 and the B.S. degree in Interdisciplinary Natural Sciences from the University of South Florida, Tampa, FL, in 2018.

From 2024 to 2025, he interned at the Johns Hopkins University Applied Physics Laboratory. Previous to this experience, he was a research intern in the Science \& Technology Directorate of the U.S. Department of Homeland Security in 2023. 
His research focuses on low‑shot learning, unseen threat detection, and deep learning applications in computer vision and cybersecurity. He is an IEEE Student Member and an affiliated student with the Institute for Human and Machine Cognition.
\end{IEEEbiography}

\begin{IEEEbiography}[{\includegraphics[width=1in,height=1.25in,clip,keepaspectratio]{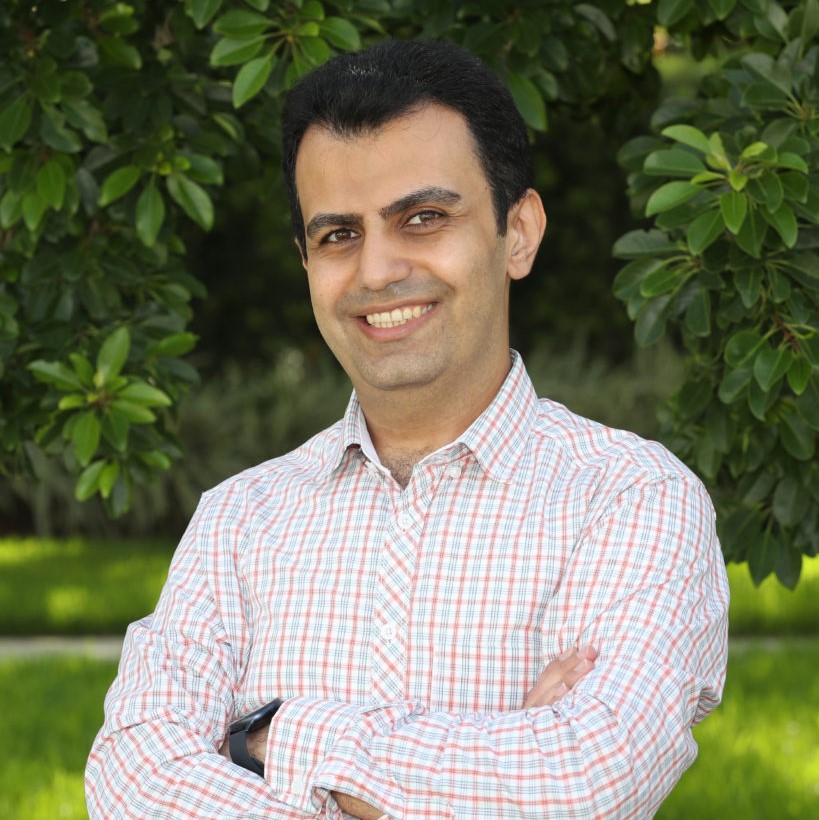}}]{Andrew A. Mahyari} received his BSc., MSc., and PhD. in Electrical Engineering and MSc in Statistics from Michigan State University in 2016. From 2017-2018, he worked at ABB Robotics R\&D Center in San Jose, CA. He is a research scientists at Florida Institute for Human and Machine Cognition and research professor at the department of Intelligent Systems and Robotics at University of West Florida.

\end{IEEEbiography}


\begin{IEEEbiography}
[{\includegraphics[width=1in,height=1.25in,clip,keepaspectratio]{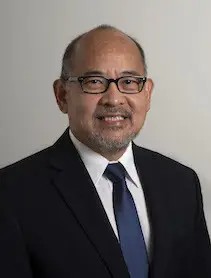}}]{GUILLERMO FRANCIA, III} received his MSc and PhD in Computer Science from New Mexico Tech. He joined the University of West Florida Center for Cybersecurity in 2018. Previously, Dr. Francia served as the Director of the Center for Information Security and Assurance and held a Distinguished Professor position at Jacksonville State University. He twice received a Fulbright scholarship award: teaching award in Malta (2007) and research award in the UK (2017). Dr. Francia is a recipient of numerous cybersecurity research and curriculum development grants. 
\end{IEEEbiography}

\begin{IEEEbiography}
[{\includegraphics[width=1in,height=1.25in,clip,keepaspectratio]{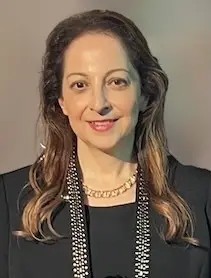}}]{Eman El-Sheikh} received her MSc and PhD in Computer Science from Michigan State University. She is professor of computer science at University of West Florida and serves as Associate Vice President and Professor at the
University of West Florida Center for Cybersecurity. Dr. El-Sheikh also serves as the Chief Strategic Alliance Officer and USA Ambassador for the Global Council for Responsible AI.
\end{IEEEbiography}




\EOD

\end{document}